\documentclass[twocolumn,english]{revtex4-2}
\usepackage{lmodern}
\usepackage{lmodern}

\usepackage[T1]{fontenc}
\usepackage[latin9]{inputenc}
\usepackage{babel}
\usepackage{amsmath}
\usepackage{amssymb}
\usepackage{graphicx}
\PassOptionsToPackage{normalem}{ulem}
\usepackage{ulem}
\usepackage[unicode=true,pdfusetitle,
 bookmarks=true,bookmarksnumbered=false,bookmarksopen=false,
 breaklinks=false,pdfborder={0 0 1},backref=false,colorlinks=false]
 {hyperref}

\makeatletter

\providecommand{\tabularnewline}{\\}

\usepackage{babel}

\makeatother

\begin{document}
\title{The number of populated electronic configurations in a hot dense plasma}
\author{Menahem Krief}
\email{menahem.krief@mail.huji.ac.il}

\address{Racah Institute of Physics, The Hebrew University, 9190401 Jerusalem,
Israel}
\begin{abstract}
In hot dense plasmas of intermediate or high-Z elements in the state
of local thermodynamic equilibrium, the number of electronic configurations
contributing to key macroscopic quantities such as the spectral opacity
and equation of state, can be enormous. In this work we present systematic
methods for the analysis of the number of relativistic electronic
configurations in a plasma. While the combinatoric number of configurations
can be huge even for mid-Z elements, the number of configurations
which have non negligible population is much lower and depends strongly
and non-trivially on temperature and density. We discuss two useful
methods for the estimation of the number of populated configurations:
(i) using an exact calculation of the total combinatoric number of
configurations within superconfigurations in a converged super-transition-array
(STA) calculation, and (ii) by using an estimate for the multidimensional
width of the probability distribution for electronic population over
bound shells, which is binomial if electron exchange and correlation
effects are neglected. These methods are analyzed, and the mechanism
which leads to the huge number of populated configurations is discussed
in detail. Comprehensive average atom finite temperature density functional
theory (DFT) calculations are performed in a wide range of temperature
and density for several low, mid and high Z plasmas. The effects of
temperature and density on the number of populated configurations
are discussed and explained. 
\end{abstract}
\maketitle

\section{Introduction}

The calculation of radiative transport properties and equations state
from first principles are of key importance in the modeling of a wide
variety of high energy density plasmas, which exist both in stellar
interiors \cite{christensen2009opacity,villante2015quantitative,krief2016line,mendoza2018computation,pain2017detailed,fontes2015relativistic,fontes2015alamos,colgan2016new,iglesias2015iron,iglesias1996updated,badnell2005updated}
and in terrestrial laboratories, such as Z-pinch and high-power laser
facilities \cite{bailey2015higher,nagayama2016calibrated,nagayama2016model,nagayama2019systematic,back2000detailed,fryer2016uncertainties,moore2015characterization,fleurot2005laser,decker1997hohlraum}.
These macroscopic quantities entail a very sophisticated interplay
between plasma physics and atomic physics.

In the calculation of spectral opacities, the bound-bound photoabsorption
spectra results from all radiative transitions between all levels
from all pairs of electronic configurations. In a hot dense plasma,
the number of lines between each pair of configurations may be enormous
(\cite{scott2010advances,gilleron2009efficient}) and statistical
methods must be used. The Unresolved-Transition-Array (UTA) method
(\cite{moszkowski1962energy,bauche1979variance,bauche1988transition,krief2015variance})
treats all levels between each pair of configurations statistically,
using analytic expressions for the energy moments of the transition
array. In many cases, the number of configurations may also be extremely
large and a such detailed-configuration-accounting (DCA) calculations
are intractable as well. Similarly, in the calculation of equations
of state, a huge number of electronic configurations should be taken
into account, for an accurate calculation of the total partition function.

In this work we discuss in detail the number of electronic configurations
in a hot dense plasma. The combinatoric number of possible configurations
as well as the number of configurations which have a non negligible
population are examined. The combinatoric number of configurations
is calculated by using exact recursive relations \cite{gilleron2004stable}.
Two methods for the estimation of the number of populated configurations
are considered and compared. The calculations are performed in a very
wide range of plasma temperature (100eV-$10$keV) and density ($10^{-3}-10^{3}\text{g/cm}^{3}$)
for various low, mid and high Z elements, using finite temperature
average atom DFT calculations as well as super-transition-array (STA)
calculations, employing the opacity code STAR \cite{krief2016solar,krief2018new}.
The effects of temperature and density on the number of populated
configurations are discussed and explained.

A relativistic electronic configuration $C$ is defined by a set of
occupation numbers $\left\{ q_{s}\right\} $ on relativistic $s=(nlj)$
orbital shells, which are full solutions of the Dirac equation. In
the statistical configuration approximation (neglecting the atomic
structure within configurations \cite{gilleron2011corrections,krief2015effect}),
the occupation of a configuration $C$ is given by the Boltzmann distribution:
\begin{equation}
P_{C}=\frac{g_{C}e^{-\left(E_{C}-\mu Q_{C}\right)/k_{B}T}}{U_{tot}},\label{eq:pc}
\end{equation}
where $Q_{C}=\sum_{s}q_{s}$ is the number of bound electrons in $C$,
$\mu$ is the chemical potential, the statistical weight of $C$ is:
\begin{equation}
g_{C}=\prod_{s}\binom{g_{s}}{q_{s}},
\end{equation}
where orbital degeneracy is $g_{s}=2j_{s}+1$, the total partition
function is: 
\begin{equation}
U_{tot}=\sum_{C}g_{C}e^{-\left(E_{C}-\mu Q_{C}\right)/k_{B}T},
\end{equation}
and the the configuration average energy is: 
\begin{equation}
E_{C}=\sum_{s}q_{s}I_{s}+\frac{1}{2}\sum_{r,s}q_{r}(q_{s}-\delta_{rs})H_{rs},\label{eq:Ecav}
\end{equation}
where, the residual energy is: 
\begin{equation}
I_{s}=\epsilon_{s}-\left\langle s\left\vert \frac{Z}{r}+V\left(r\right)\right\vert s\right\rangle ,
\end{equation}
where $\epsilon_{s}$ is the orbital energy, $V\left(r\right)$ is
the self-consistent mean field central atomic potential, and $H_{rs}$
is the average interaction energy of the two-electron configuration
$rs$, given in terms of relativistic direct and exchange Slater integrals
(explicit expression can be found in Refs. \cite{bar1995effect,ovechkin2014reseos,krief2015variance}).

\section{Combinatoric number of configurations\label{sec:Combinatoric-number-of}}

First we demonstrate how to calculate the combinatoric number of configurations
with $Q$ electrons, which are distributed over a set of shells $A=\left\{ s_{1}s_{2}...s_{N}\right\} $,
defined by: 
\begin{eqnarray*}
\mathcal{N}_{Q}^{A} & = & \sum_{\substack{\{q_{s}\}_{s=1}^{N}\\
\text{with }\sum_{s=1}^{N}q_{s}=Q
}
}1,
\end{eqnarray*}
where we have expressed the constraint that only configurations $\{q_{s}\}_{s=1}^{N}$
with a total number of $Q$ electrons are summed. This sum can be
expressed in terms of partials sums over the population of the $N$th
shell:

\begin{equation}
\mathcal{N}_{Q}^{A}=\sum_{q_{N}=0}^{\text{min}(g_{N},Q)}\sum_{\substack{\{q_{s}\}_{s=1}^{N-1}\\
\text{with }\sum_{s=1}^{N-1}q_{s}=Q-q_{N}
}
}1.
\end{equation}
Since the inner sum is actually $\mathcal{N}_{Q-q_{N}}^{A/\{s_{N}\}}$,
where $A/\{s_{N}\}$ denotes the set $A$ excluding the shell $s_{N}$,
we get the recursive relation: 
\begin{eqnarray}
\mathcal{N}_{Q}^{A} & = & \sum_{q_{N}=0}^{\text{min}(g_{N},Q)}\mathcal{N}_{Q-q_{N}}^{A/\{s_{N}\}}.\label{eq:rec_c}
\end{eqnarray}
We note that this recursion relation is performed simultaneously over
the numbers of electrons $Q$ and the number of shells $N$, and should
obey the initial condition $\mathcal{N}_{Q}^{A}=\delta_{Q,0}$ for
an empty shell group $A$. The derivation above, which results from
a simple combinatoric argument, can be proved by employing the powerful
and general method of generating functions, as was done in detail
Ref. \cite{gilleron2004stable}, which deals with stable algorithms
for the calculation of canonical partition functions, which were further
generalized and improved in Refs. \cite{wilson2007further,pain2020optimized}.
We note that in a recent work \cite{pain2020analytical}, a novel
approach, which leads to new recurrence and analytic relations as
well as a novel statistical modeling of the combinatoric number of
configurations is developed in detail.

The total combinatoric number of configurations for an element with
an atomic number $Z$, is given by summing the number of configurations
over all ionization levels: 
\begin{equation}
\mathcal{N}_{C}^{\text{combin}}=\sum_{Q=0}^{Z}\mathcal{N}_{Q}^{A},\label{eq:NC_COMBIN}
\end{equation}
where $A$ is the set of all shells from which configurations are
constructed. We note that, in general, the number of bound shells
is a function of the atomic number, temperature and density, which
determines the self consistent central potential - for example, a
higher Z element has a larger number of bound shells, due to the higher
nucleus charge and a higher density plasma may have a smaller number
of shells due to an increased pressure ionization effect (bound states
dissolving into the continuum \cite{liberman1979self,Rozsnyai1972,blenski1995pressure,wilson2006purgatorio,Novikov2011,ovechkin2014reseos,ovechkin2019plasma,krief_dh_2018_apj,gill2017tartarus,starrett2019wide}).

One can also define the number of configurations taking into account
only charge states with probability larger than $p$:
\begin{equation}
\mathcal{N}_{C}^{\text{combin}}\left(p\right)=\sum_{Q,\ P_{Q}>p}\mathcal{N}_{Q}^{A},\label{eq:NC_COMBIN_P}
\end{equation}
where the charge state distribution is given by the sum of probabilities
of all configurations with a total charge $Q$:
\begin{equation}
P_{Q}=\sum_{C,\ Q_{C}=Q}P_{C}.\label{eq:pc_C}
\end{equation}
We note that eq. \ref{eq:NC_COMBIN_P} may depend strongly on the
parameter $p$, and may give a severe overestimation for the number
of \uline{populated} configurations, since each charge state $Q$
can correspond to a huge number of configurations, some of which have
an extremely low probability. In addition, an exact computation of
the ion charge distribution \ref{eq:pc_C} (via eq. \ref{eq:pc})
is in many cases intractable, due the huge number of configurations
that need to be taken into account. However, many approximated methods
for the calculation of the charge state distribution in a plasma exist
(for example, via the well known Saha equations) and can be used ``externally''
in the calculation of eq. \ref{eq:NC_COMBIN_P}. In this work we will
use a more accurate charge state distribution which is obtained from
an STA calculation (see below).

\section{Number configurations within superconfigurations\label{sec:Number-configurations-within}}

In the STA method \cite{BarShalom1989,blenski2000superconfiguration,hazak2012configurationally,kurzweil2016summation,ovechkin2014reseos,wilson2015partially,krief2016solar,krief2018star,bauche2015atomic,pain2021super},
a large number of configurations $C$ are grouped into super-configurations
(SCs), commonly denoted by $\Xi=\Pi_{\sigma}\sigma^{Q_{\sigma}}$,
which are defined as sets of configurations which have $Q_{\sigma}$
electrons in each 'supershell' $\sigma$, which is a group of shells.
The total occupation of a super configuration is naturally:

\begin{equation}
P_{\Xi}=\sum_{C\in\Xi}P_{C},
\end{equation}
and combinatoric number of configurations within a superconfiguration
is:

\begin{equation}
\mathcal{N}_{C}\left(\Xi\right)=\sum_{C\in\Xi}1.\label{eq:nsc}
\end{equation}
As was already noted in Ref. \cite{gilleron2004stable}, $\mathcal{N}_{\Xi}$
can be evaluated exactly, using the recursive relation \ref{eq:rec_c}.
Since electron occupation numbers in different supershells are independent,
eq. \ref{eq:rec_c} can be used for each supershell, to give the combinatoric
number of\textbf{ }configurations withing a superconfiguration: 
\begin{equation}
\mathcal{N}_{C}\left(\Xi\right)=\prod\limits _{\sigma}\mathcal{N}_{Q_{\sigma}}^{\sigma},
\end{equation}
where $\mathcal{N}_{Q_{\sigma}}^{\sigma}$ is calculated by applying
the recursion relation \ref{eq:rec_c} for each supershell.

A simple estimate for the number of populated configurations, is given
by the number of configurations within all populated superconfigurations
with a probability larger than $p$:

\begin{equation}
\mathcal{N}_{C}^{\text{in SCs}}\left(p\right)=\sum_{\substack{\Xi\\
P_{\Xi}>p
}
}\mathcal{N}_{C}\left(\Xi\right).\label{eq:nc_insc}
\end{equation}
where we have expressed the constraint that only superconfigurations
with a non-negligible occupation are included.

We note that the combinatoric number of superconfigurations, for a
given set of supershells, can also be calculated using the recursion
relation \ref{eq:rec_c}, by treating the supershells $\sigma$ as
shells - defining $A=\left\{ \sigma_{1}...\sigma_{N}\right\} $, with
total degeneracies $g_{\sigma}=\sum_{s\in\sigma}g_{s}$ and summing
over all ionization levels.

\section{Estimation of the number of populated configuration}

\begin{figure}
\begin{centering}
\includegraphics[scale=0.065]{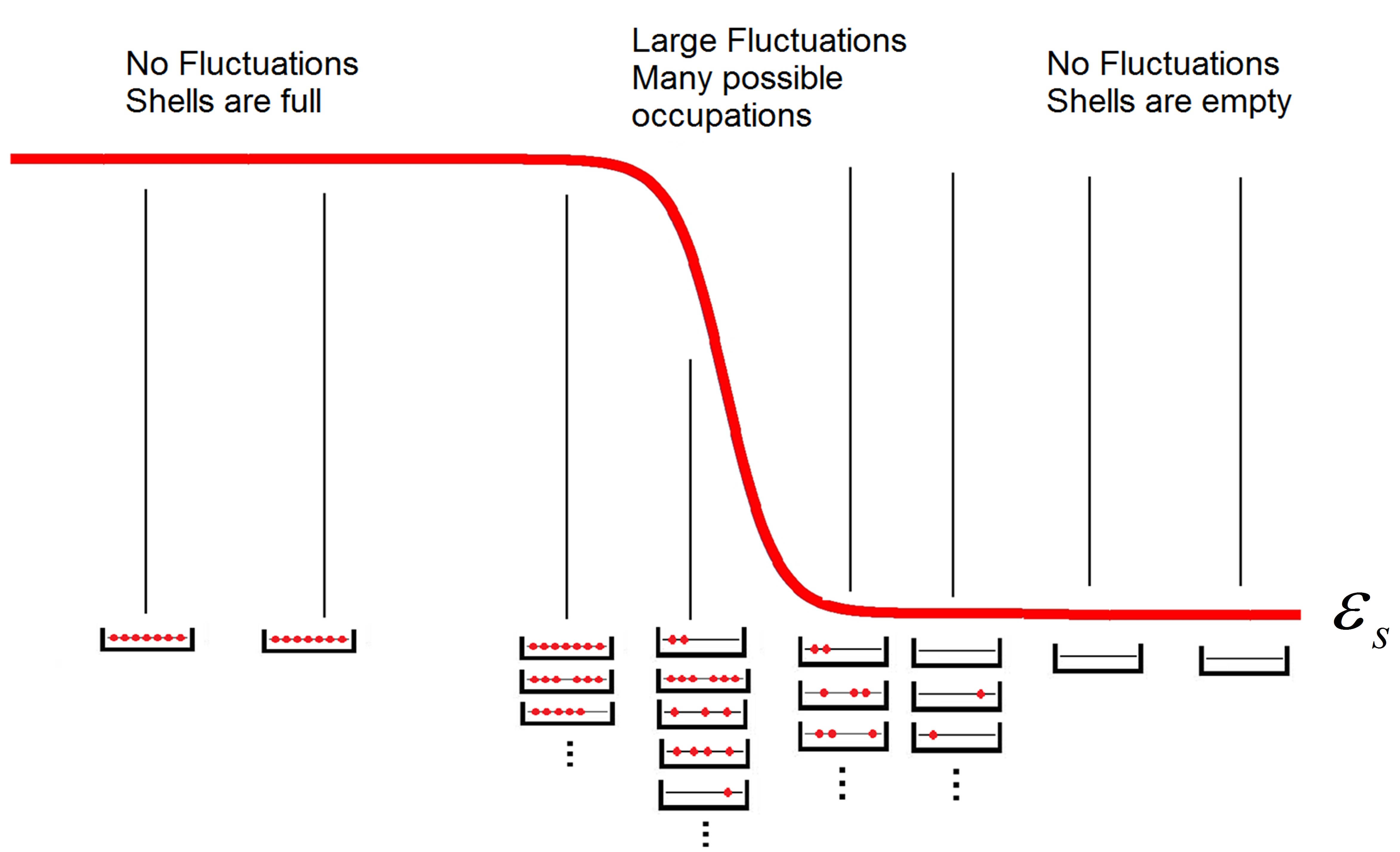} 
\par\end{centering}
\caption{(Color online) A schematic description of the population of electrons over shells
according to the Fermi-Dirac distribution (red thick line). The occupation of electronic
shells which have energies nearby the Fermi-Dirac step (which is located
at $\epsilon_{s}\approx\mu$ and has a width of the order of $k_{B}T$),
fluctuate and result in wide range of possibilities to distribute
electrons over the magnetic quantum numbers, while shells which are
far from the Fermi-Dirac step are either filled or empty. The fluctuating
occupation numbers may give rise the a huge number of populated electronic
configurations. \label{fig:fd_illustration}}
\end{figure}

\begin{table}
\begin{centering}
\begin{tabular}{|c|c|c|}
\hline 
supershell $\sigma$  & $Q_{\sigma}$ range  & degeneracy $g_{\sigma}$\tabularnewline
\hline 
(1s)  & {[}2,2{]}  & 2\tabularnewline
\hline 
(2s)  & {[}0,2{]}  & 2\tabularnewline
\hline 
$(2\text{p}_{-})$  & {[}0,2{]}  & 2\tabularnewline
\hline 
$(2\text{p}_{+})$  & {[}0,4{]}  & 4\tabularnewline
\hline 
$(3\text{s}...5\text{f}_{-})$  & {[}0,7{]}  & 74\tabularnewline
\hline 
$(5\text{f}_{+}...8\text{i}_{+})$  & {[}0,7{]}  & 294\tabularnewline
\hline 
\end{tabular}
\par\end{centering}
\caption{The converged relativistic supershell structure (left column) for
Iron at $T=182\text{eV}$, $\rho=0.13\text{g/cm}^{3}$. The range
for the number of electrons in each supershell (middle column), chosen
such that the superconfiguration occupation is larger than $10^{-7}$,
as well as the total supershell degeneracy (right column), are also
given. The total number of superconfigurations is $\mathcal{N}_{\Xi}=2880$.
\label{tab:iron}}
\end{table}

\begin{table}
\begin{centering}
\begin{tabular}{|c|c|c|}
\hline 
supershell $\sigma$  & $Q_{\sigma}$ range  & degeneracy $g_{\sigma}$\tabularnewline
\hline 
(1s)  & {[}2,2{]}  & 2\tabularnewline
\hline 
(2s)  & {[}2,2{]}  & 2\tabularnewline
\hline 
$(2\text{p}_{-})$  & {[}2,2{]}  & 2\tabularnewline
\hline 
$(2\text{p}_{+})$  & {[}4,4{]}  & 4\tabularnewline
\hline 
$(3\text{s}3\text{p}_{-}3\text{p}_{+})$  & {[}7,8{]}  & 8\tabularnewline
\hline 
$(3\text{d}_{-}3\text{d}_{+})$  & {[}9,10{]}  & 10\tabularnewline
\hline 
$(4\text{s}...4\text{f}_{+})$  & {[}3,27{]}  & 32\tabularnewline
\hline 
$(5\text{s}...7\text{d}_{+})$  & {[}0,12{]}  & 140\tabularnewline
\hline 
$(7\text{f}_{-}...8\text{k}_{+})$  & {[}0,7{]}  & 208\tabularnewline
\hline 
\end{tabular}
\par\end{centering}
\caption{Same as table \ref{tab:iron}, for Gold at $T=200\text{eV}$, $\rho=0.1\text{g/cm}^{3}$.
The total number of superconfigurations is $\mathcal{N}_{\Xi}=10400$.\label{tab:gold}}
\end{table}

\begin{figure}
\begin{centering}
\includegraphics[scale=0.5]{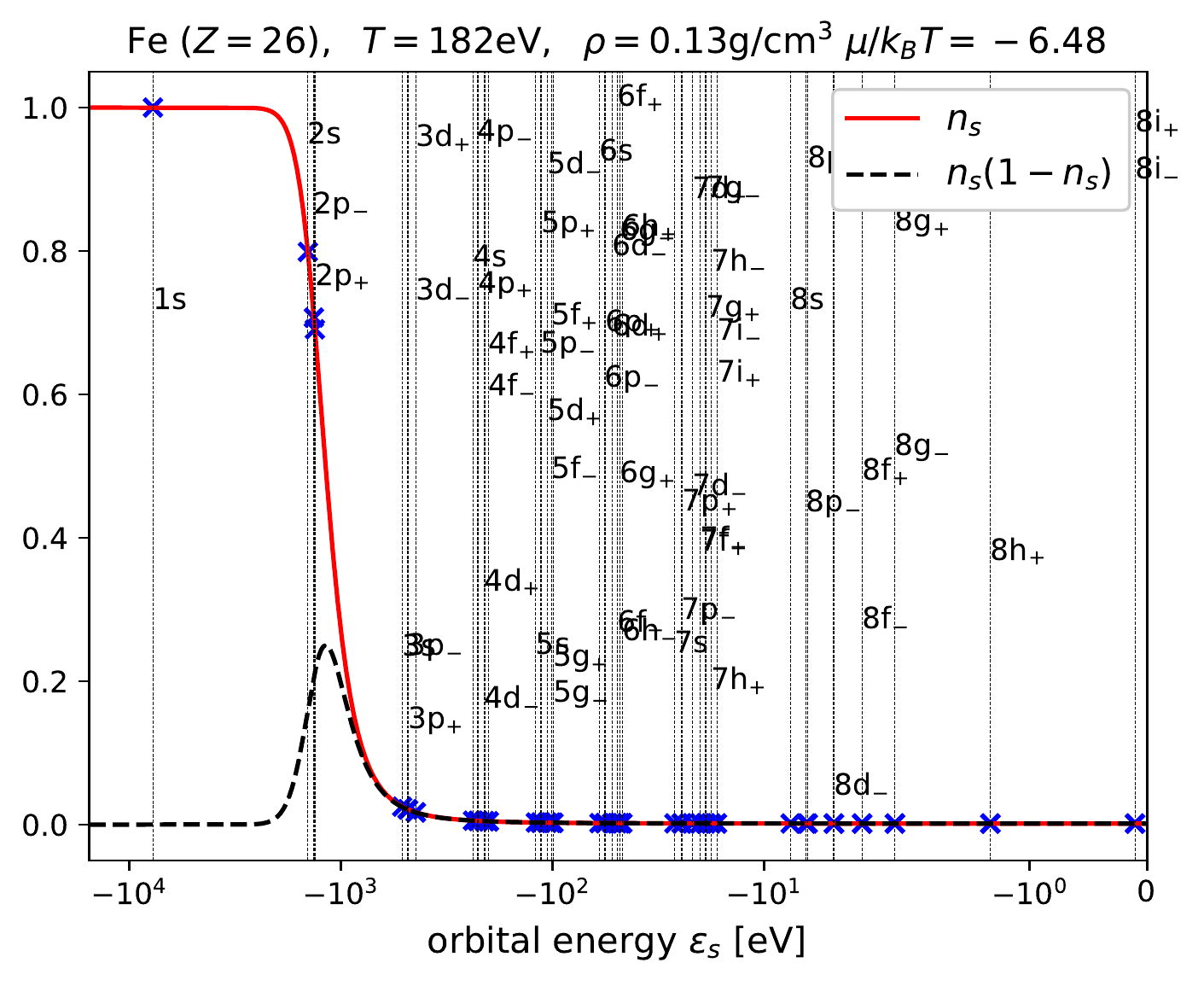} 
\par\end{centering}
\begin{centering}
\includegraphics[scale=0.5]{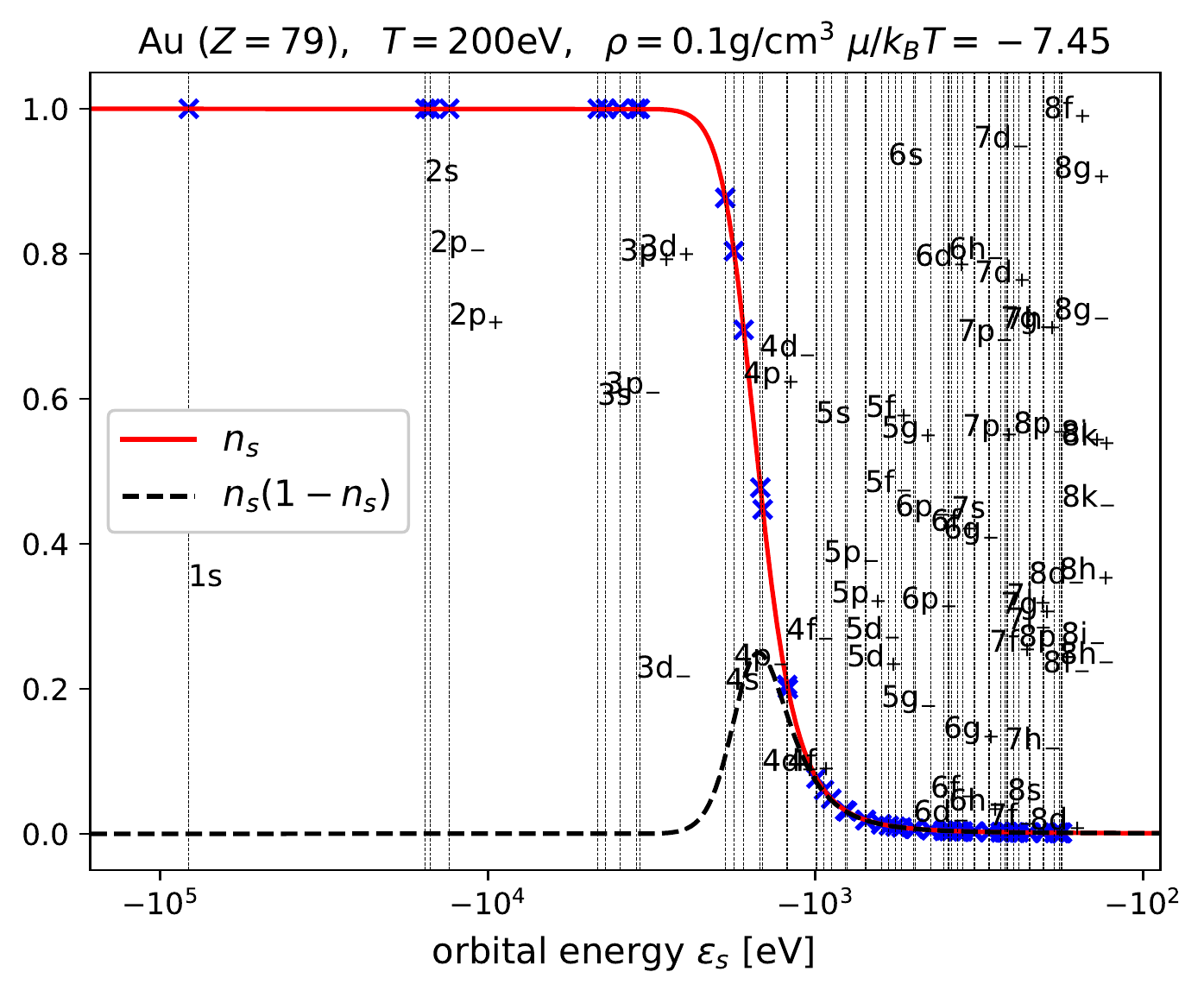} 
\par\end{centering}
\caption{(Color online) The Fermi-Dirac distribution (red solid lines, as illustrated in Fig. \ref{fig:fd_illustration})
and the occupation fluctuation $\delta q_{s}^{2}/g_{s}$ (black dashed lines, see eq. \ref{eq:fluc})) as a function of orbital energy,
for Iron at $T=182\text{eV}$, $\rho=0.13\text{g/cm}^{3}$ (upper
figure) and Gold at $T=200\text{eV}$, $\rho=0.1\text{g/cm}^{3}$
(lower figure). The relativistic Dirac bound orbitals are listed and
represented as vertical dashed thin lines at their appropriate bound energies.
The resulting chemical potential is given in the title.\label{fig:fd_fluc_fe_gold}}
\end{figure}

\begin{figure}
\begin{centering}
\includegraphics[scale=0.5]{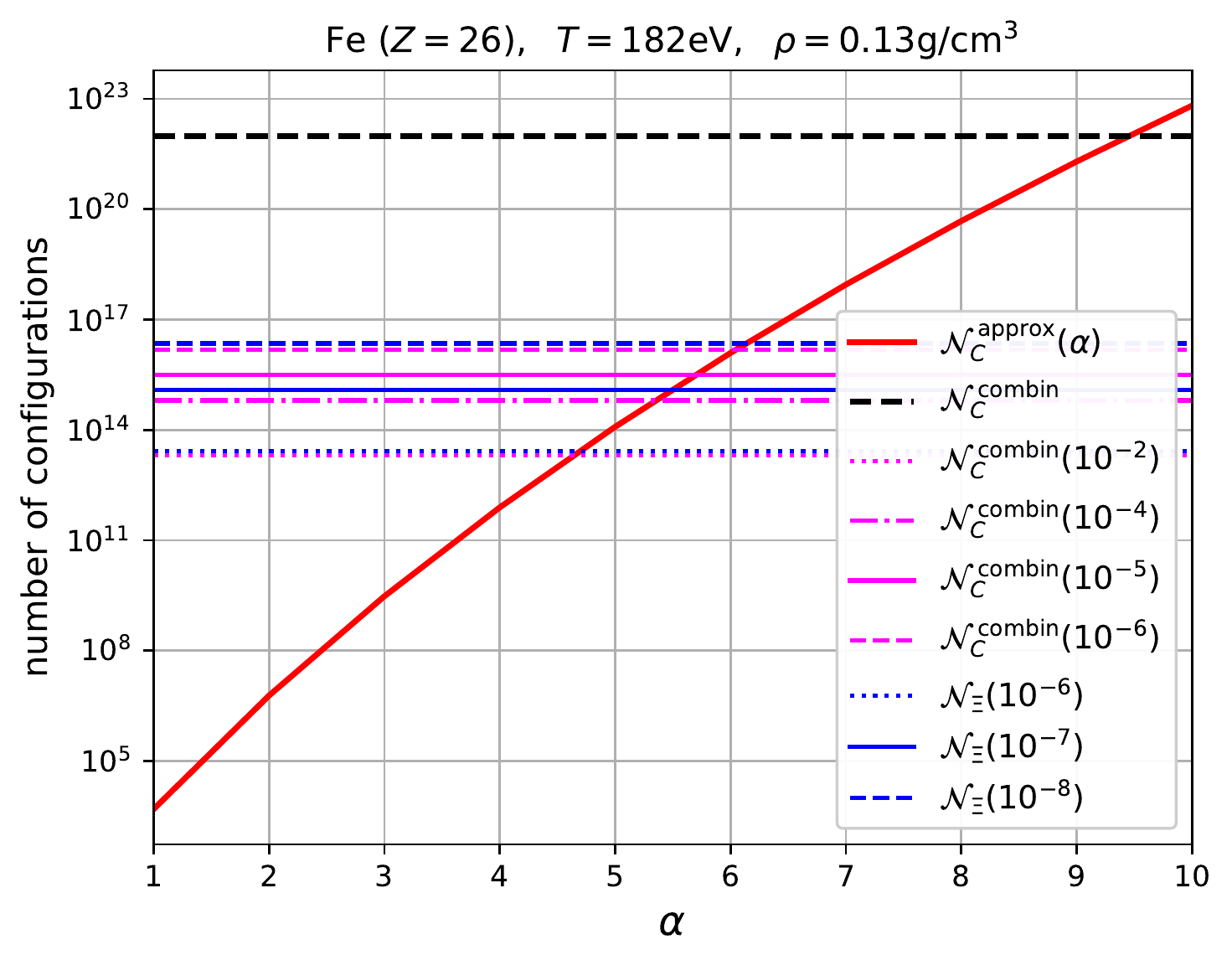} 
\par\end{centering}
\begin{centering}
\includegraphics[scale=0.5]{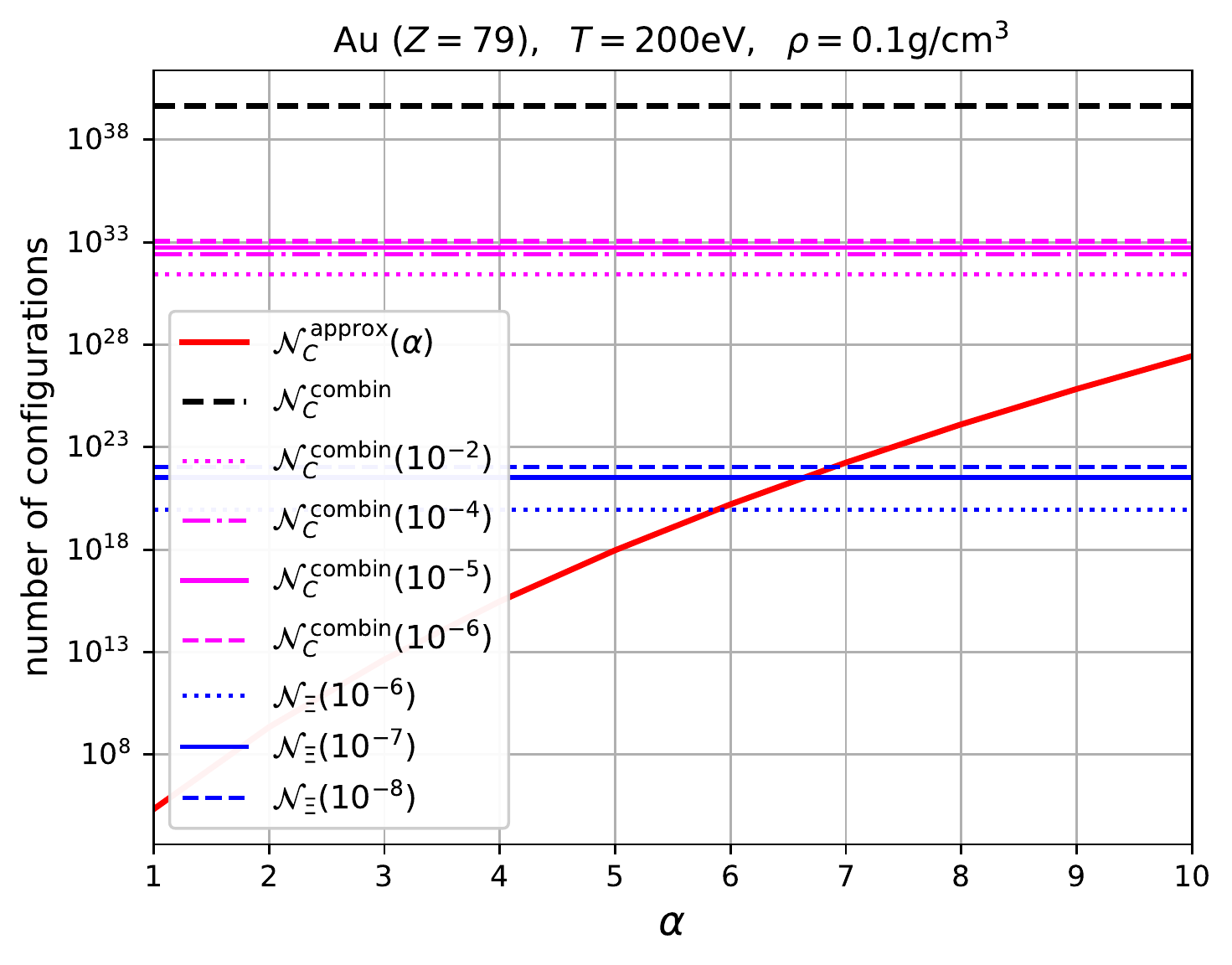} 
\par\end{centering}
\caption{(Color online) The estimate given in eq. \ref{eq:num_config_binomial} for the number
of populated configurations, as a function of the parameter $\alpha$
(red solid lines), for Iron at $T=182\text{eV}$, $\rho=0.13\text{g/cm}^{3}$
(upper figure) and Gold at $T=200\text{eV}$, $\rho=0.1\text{g/cm}^{3}$
(lower figure). Also shown are the combinatoric number of configurations
over all ionization levels (eq. \ref{eq:NC_COMBIN}, black dashed lines) and
over all ionization levels with probabilities larger than $10^{-2},10^{-4},10^{-5}$
and $10^{-6}$ respectively (eq. \ref{eq:NC_COMBIN_P}, magenta lines),
as well as the number of configurations within populated superconfigurations
(eq. \ref{eq:nc_insc}), with probabilities larger than $10^{-6},10^{-7}$
and $10^{-8}$, respectively (eq. \ref{eq:nc_insc}, blue lines).\label{fig:alpha}}
\end{figure}

\begin{figure}
\begin{centering}
\includegraphics[scale=0.5]{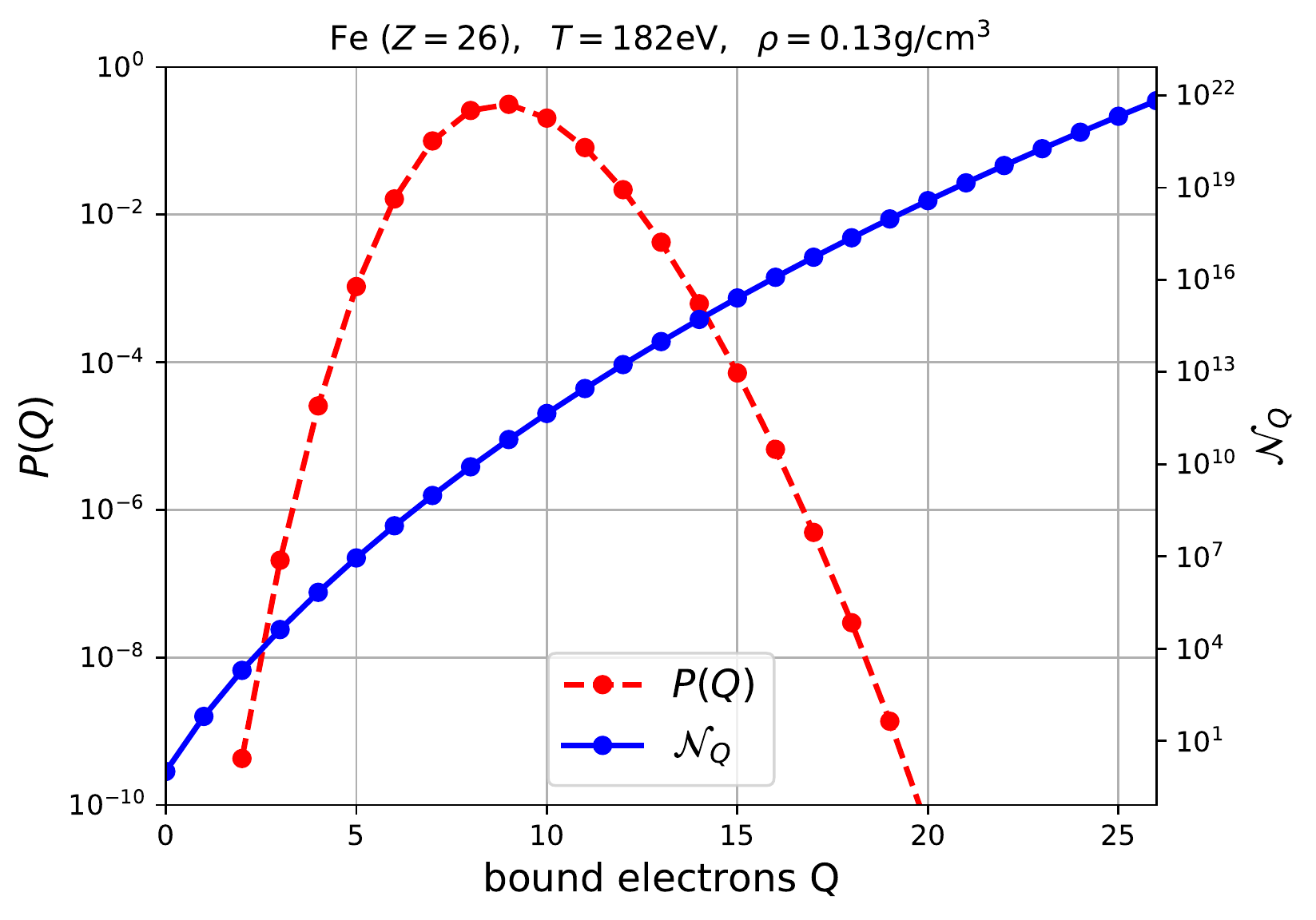}
\par\end{centering}
\begin{centering}
\includegraphics[scale=0.5]{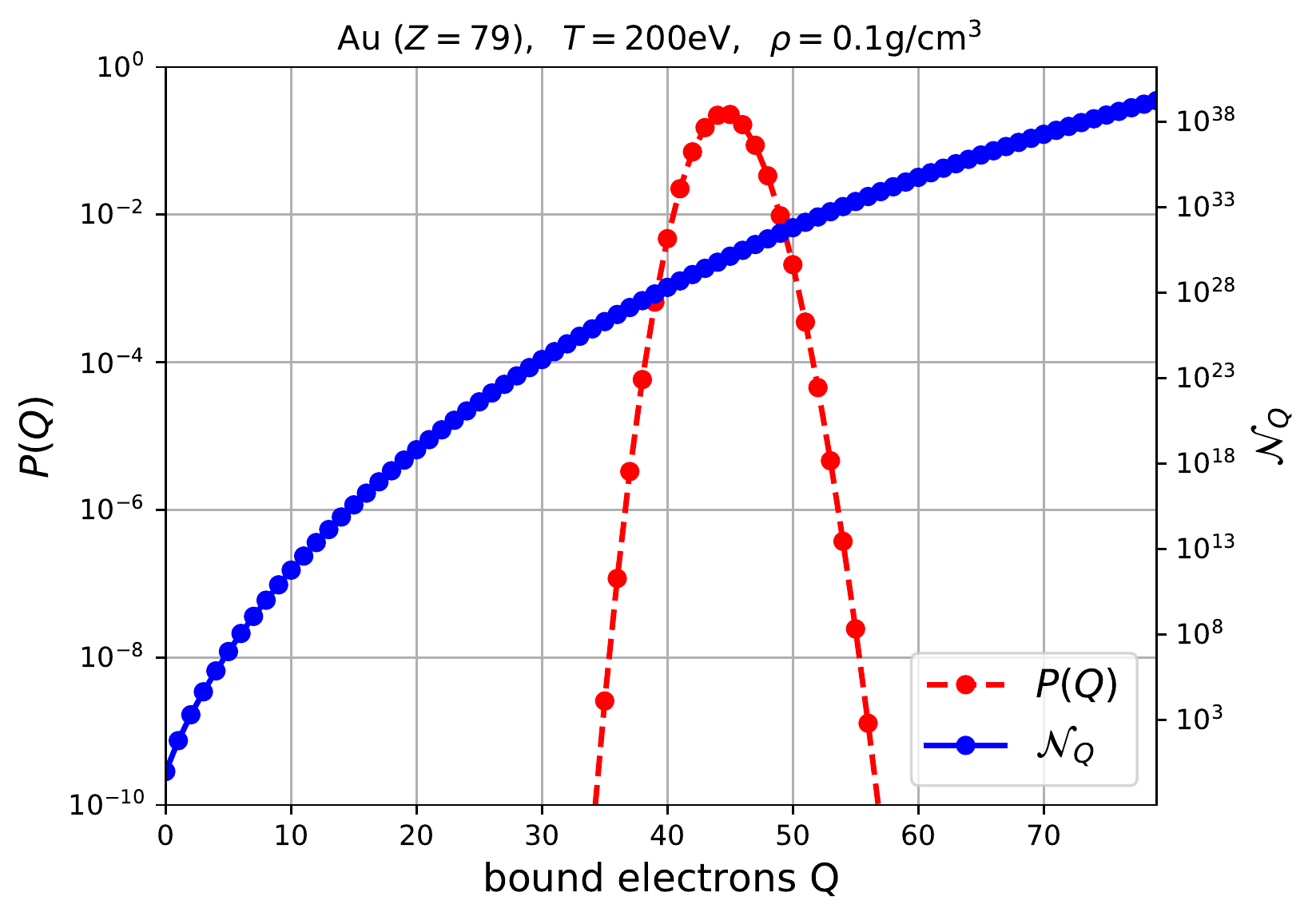}
\par\end{centering}
\caption{(Color online) The charge state distribution calculated using an STA model (red dashed lines,
left axis) and the combinatoric number of configurations for each
charge state (eq. \ref{eq:NC_COMBIN}, blue lines, on the right axis),
for Iron at $T=182\text{eV}$, $\rho=0.13\text{g/cm}^{3}$ (upper
figure) and Gold at $T=200\text{eV}$, $\rho=0.1\text{g/cm}^{3}$
(lower figure).\label{fig:pq}}
\end{figure}

In this section we discuss a very simple way to estimate the number
of populated configurations, without using a state of the art sophisticated
STA method, as was suggested in the previous section. This method
was presented in the seminal book \cite{nikiforov2006quantum} and
was used subsequently in Refs. \cite{krief2016solar,krief2018new}.
In this method, only the bound shells and chemical potential are needed
- so that the estimate for the number of populated configurations
can be obtained, for example, by solving the Dirac equation in a Thomas-Fermi
potential \cite{feynman1949equations} or in a more advanced average
atom model potential \cite{liberman1979self,Rozsnyai1972,blenski1995pressure,Novikov2011,wilson2006purgatorio,ovechkin2014reseos,ovechkin2019plasma,krief_dh_2018_apj,gill2017tartarus,starrett2019wide}.

Let us estimate the number of populated configurations. Neglecting
the electron-electron interaction effects beyond the self consistent
field model, the configuration average energy given in eq. \ref{eq:Ecav}
can be approximated as a first order polynomial of the occupation
numbers: 
\begin{equation}
E_{C}\approx\sum_{s}q_{s}\epsilon_{s}.
\end{equation}
Under this approximation, the total partition function is: 
\begin{equation}
U_{tot}=\prod_{s}\left(1+e^{-(\epsilon_{s}-\mu)/k_{B}T}\right)^{g_{s}},
\end{equation}
and the configuration probability \ref{eq:pc} becomes a simple multivariate
binomial distribution of the occupation numbers $\left\lbrace q_{s}\right\rbrace $,
given by: 
\begin{eqnarray}
P\left(\left\{ q_{s}\right\} \right)\equiv P_{C} & = & \prod_{s}\binom{g_{s}}{q_{s}}n_{s}^{q_{s}}\left(1-n_{s}\right){}^{g_{s}-q_{s}},\label{eq:PC_BINOMIAL}
\end{eqnarray}
where $n_{s}=1/\left(e^{(\epsilon_{s}-\mu)/k_{B}T}+1\right)$ is the
Fermi-Dirac distribution. The variance of the population of each shell
is given by: 
\begin{eqnarray}
\delta q_{s} & \equiv\sqrt{\left\langle \left(q_{s}-\left\langle q_{s}\right\rangle \right){}^{2}\right\rangle } & =\sqrt{g_{s}n_{s}\left(1-n_{s}\right)}.\label{eq:fluc}
\end{eqnarray}
As illustrated in Figs. \ref{fig:fd_illustration}-\ref{fig:fd_fluc_fe_gold},
it is evident that the occupation of shells whose energies are near
the Fermi-Dirac step, which is located around $-k_{B}T\lesssim\epsilon_{s}-\mu\lesssim k_{B}T$
fluctuate, while the other shells are either filled or empty. These
fluctuating shells may have a wide range of possibilities to distribute
electrons over the magnetic quantum numbers $0\leq m_{s}\leq g_{s}$.
The fluctuating occupation numbers may give rise the a huge number
of populated configurations, which increases exponentially with the
number of fluctuating shells.

The number of populated configurations can be estimated as the number
of possibilities to put electrons in each shell, within a few standard
deviations $\delta q_{s}$ around the average occupation $\left\langle q_{s}\right\rangle =g_{s}n_{s}$
of the multivariate distribution \ref{eq:PC_BINOMIAL}. The number
of possible occupation numbers for each shell is estimated as $\alpha\times\delta q_{s}$,
where $\alpha/2$ is the number of standard deviations. Therefore,
the number of populated configurations can be estimated by: 
\begin{eqnarray}
\mathcal{N}_{C}^{\text{approx}} & = & \prod_{s}\left(\alpha\delta q_{s}+1\right),\label{eq:num_config_binomial}
\end{eqnarray}
which is simply the multidimensional ``width'' of the multivariate
binomial distribution (\ref{eq:PC_BINOMIAL}).

We note that the result may depend on the somewhat arbitrary value
chosen for $\alpha$, but it can be expected that a reasonable value
should be in the range $2\lesssim\alpha/2\lesssim4$ - corresponding
to a range of two to four standard deviations for the occupation of
each shell. In order to demonstrate this, $\mathcal{N}_{C}^{\text{approx}}$
was calculated as a function of $\alpha$ in the range $1\leq\alpha\leq10$
and compared with the number of configurations within populated superconfigurations
(eq. \ref{eq:nc_insc}) of a converged STA calculation. Two cases
are considered: (1) Iron (Z=26) at typical conditions of the recent
Sandia Z experiments \cite{bailey2015higher,nagayama2019systematic},
with temperature $T=182\text{eV}$ and density $\rho=0.13\text{g/cm}^{3}$
and (2) Gold (Z=79) with temperature $T=200\text{eV}$ and density
$\rho=0.1\text{g/cm}^{3}$. The calculations were performed using
the relativistic average-atom model implemented in the STA code STAR
\cite{krief2015effect,krief2015variance,krief2016line,krief2016solar,krief2018new,krief2018star,krief_dh_2018_apj}.
The number of bound shells was limited to a principle atomic number
of $n_{\text{max}}=8$ (which corresponds to 64 relativistic orbitals),
since highly excited bound orbitals can be accounted for by using
the method detailed in Ref. \cite{pain2015accounting}, and therefore
need not be accounted in the estimation of the number of populated
configurations that are used in the calculation of spectral opacities.
Illustration of the bound orbitals and the Fermi-Dirac step for the
two cases is given Fig. \ref{fig:fd_fluc_fe_gold}. The supershell
structure for each case are given in tables \ref{tab:iron}-\ref{tab:gold}.
In Fig. \ref{fig:alpha} we present the number of populated configurations
as a function of $\alpha$, in comparison to the combinatoric number
of configurations over all ionization levels (eq. \ref{eq:NC_COMBIN})
and over all ionization levels with occupation probabilities larger
than $10^{-2},10^{-4},10^{-5}$ and $10^{-6}$ (eq. \ref{eq:NC_COMBIN_P})
as well as the number of configurations within populated superconfigurations
(eq. \ref{eq:nc_insc}), with occupation probabilities larger than
$10^{-6},10^{-7}$ and $10^{-8}$. These results are also given explicitly
in Table \ref{tab:numc}. The strong dependence on $\alpha$ is evident,
and it seems, as expected, that a good choice is $\alpha=6$, which
corresponds to $3$ standard deviations for the occupation of each
shell. 

It is evident that the probability thresholds affect the resulting
number of configurations by about 1-2 orders of magnitude, which is
a reasonable accuracy for the number of configurations, which can
be of the order of $10^{15}-10^{40}$. It is evident that for the
Iron case, the results for $\mathcal{N}_{C}^{\text{combin}}\left(p\right)$
and $\mathcal{N}_{C}^{\text{in SCs}}\left(p\right)$ agree to within
1-2 orders of magnitude. However, it is evident that for the gold
case, the $\mathcal{N}_{C}^{\text{combin}}\left(p\right)$ values
give a severe overestimation (by about ten orders of magnitude) compared
to $\mathcal{N}_{C}^{\text{in SCs}}\left(p\right)$, which represents
the correct estimate for the number of configurations that need to
be taken into account in opacity calculations. As mentioned in section
\ref{sec:Combinatoric-number-of}, this is to be expected since each
charge state $Q$ can correspond to a huge number of configurations,
some of which have extremely low probabilities. This is more likely
to happen for a high Z element, for which the supershell structure
(see table \ref{tab:gold}) gives rise to only a small fraction of
all possible configurations for some charge states. In this way $\mathcal{N}_{C}^{\text{in SCs}}$
only accounts for configurations with non-negligible probabilities
while $\mathcal{N}_{C}^{\text{combin}}\left(p\right)$ accounts for
all configurations for non-negligible charge states, without taking
into account the confrontational structure (which determines the configuration
probability in eq. \ref{eq:pc}).

\begin{table}
\begin{centering}
\begin{tabular}{|c|c|c|}
\hline 
 & $\substack{\text{Fe, }T=182\text{eV},\rho=0.13\text{g/cm}^{3}\\
\\
}
$ & $\substack{\text{Au, }T=200\text{eV, \ensuremath{\rho}=0.1\ensuremath{\text{g/cm}^{3}}}\\
\\
}
$\tabularnewline
\hline 
$\mathcal{N}_{C}^{\text{combin}}$ & $9.78\times10^{21}$ & $4.64\times10^{39}$\tabularnewline
\hline 
$\mathcal{N}_{C}^{\text{combin}}\left(10^{-2}\right)$ & $2.08\times10^{13}$ & $2.71\times10^{31}$\tabularnewline
\hline 
$\mathcal{N}_{C}^{\text{combin}}\left(10^{-4}\right)$ & $6.34\times10^{14}$ & $2.6\times10^{32}$\tabularnewline
\hline 
$\mathcal{N}_{C}^{\text{combin}}\left(10^{-5}\right)$ & $3.2\times10^{15}$ & $5.39\times10^{32}$\tabularnewline
\hline 
$\mathcal{N}_{C}^{\text{combin}}\left(10^{-6}\right)$ & $1.53\times10^{16}$ & $1.1\times10^{33}$\tabularnewline
\hline 
$\mathcal{N}_{C}^{\text{in SCs}}\left(10^{-6}\right)$ & $2.71\times10^{13}$ & $9.01\times10^{19}$\tabularnewline
\hline 
$\mathcal{N}_{C}^{\text{in SCs}}\left(10^{-7}\right)$ & $1.22\times10^{15}$ & $3.35\times10^{21}$\tabularnewline
\hline 
$\mathcal{N}_{C}^{\text{in SCs}}\left(10^{-8}\right)$ & $2.26\times10^{16}$ & $1.1\times10^{22}$\tabularnewline
\hline 
$\mathcal{N}_{C}^{\text{approx}}$$\left(\alpha=6\right)$ & $1.25\times10^{16}$ & $1.63\times10^{20}$\tabularnewline
\hline 
\end{tabular}
\par\end{centering}
\caption{Various values for the number of configurations for the two cases
shown in Fig. \ref{fig:alpha}.\label{tab:numc}}
\end{table}

Fig. \ref{fig:pq} shows the charge state distributions, which were
obtained from converged STA calculations, together with the combinatoric
number of configurations as a function of the number of bound electrons
(eq. \ref{eq:rec_c}), for the Iron and Gold cases. The exponential
growth for the combinatoric number of configurations for large values
of bound electrons is evident, as expected. We note that the number
of configurations in eq. \ref{eq:NC_COMBIN_P} is obtained by summing
the number of configurations per charge state in Fig. \ref{fig:pq}
in the overlapping range with the charge state distribution. 

Finally, we note that it is possible to overcome the binomial approximation
\ref{eq:PC_BINOMIAL} using the correlated probability formalism.
In Ref. \cite{green1964statistical} electron-electron interactions
are taken into account in the screening constant model which is applicable
for small interactions. The resulting correlation coefficients are
calculated to second order in the interaction energy. As noted by
Perrot and Blenski in Ref. \cite{perrot2000electronic}, this method
is complicated to implement in practice and performs poorly for low
temperatures. Perrot and Blenski introduced in Ref. \cite{perrot2000electronic}
a simple method, which overcomes these difficulties by replacing the
binomial distribution with its correlated Gaussian continuous limit,
which is applicable for orbital shells with large degeneracy and which
are not close to being full or empty. As was shown in Ref. \cite{wilson1993evaluating},
this approximation is accurate to less than 1\%. Hence, by using this
Gaussian approximation in order to calculate the population variance
of shells with a large degeneracy and which are not close to being
full or empty, and the binomial approximation for the remaining shells
(as was done above), can give a better approximation for the number
of configurations, which takes into account electron-electron interactions.
However, this is beyond the scope of this manuscript, and will be
performed in a future work.

\section{Results\label{sec:Results}}

\begin{figure*}
\begin{centering}
\includegraphics[scale=0.28]{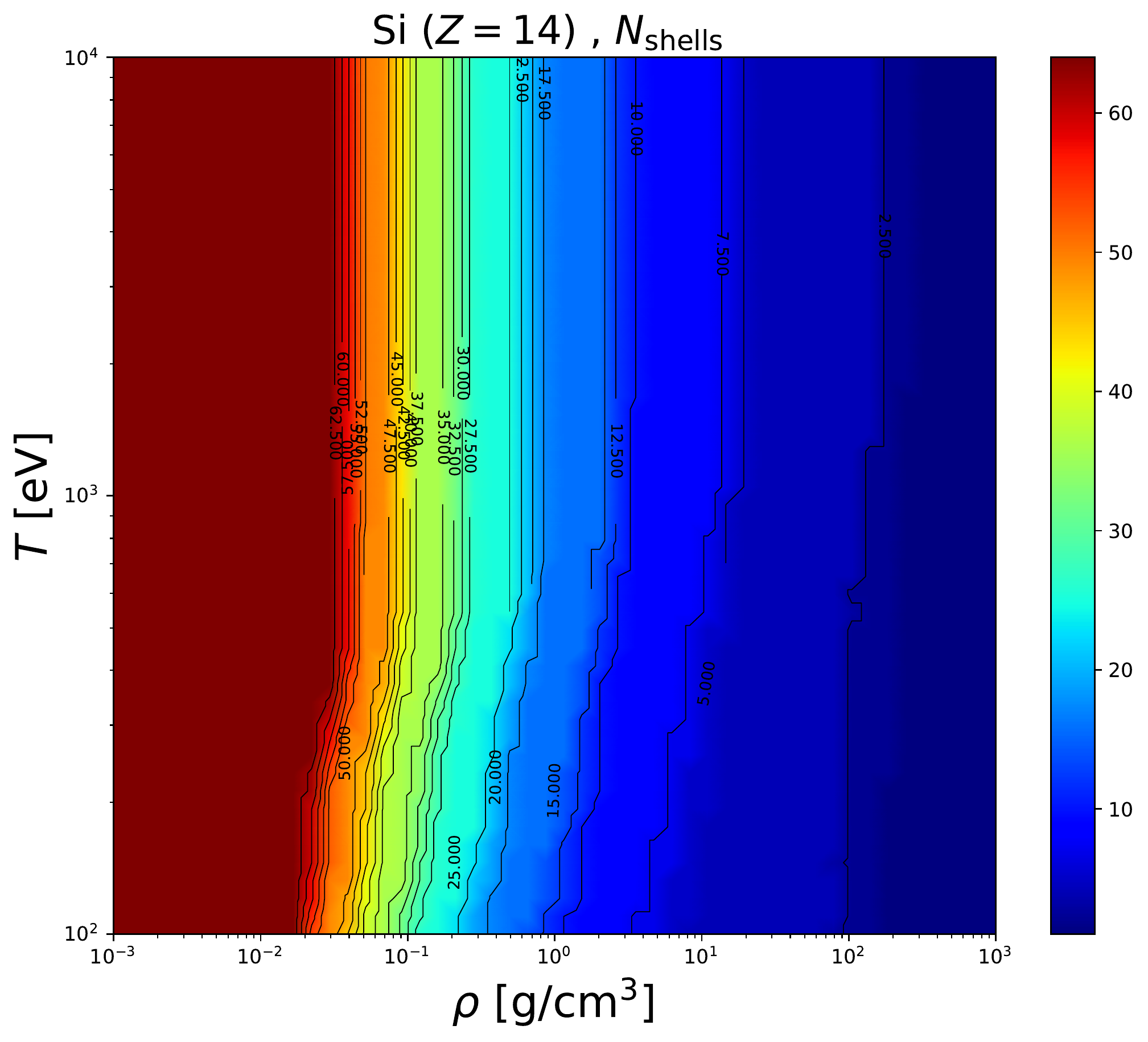}\includegraphics[scale=0.28]{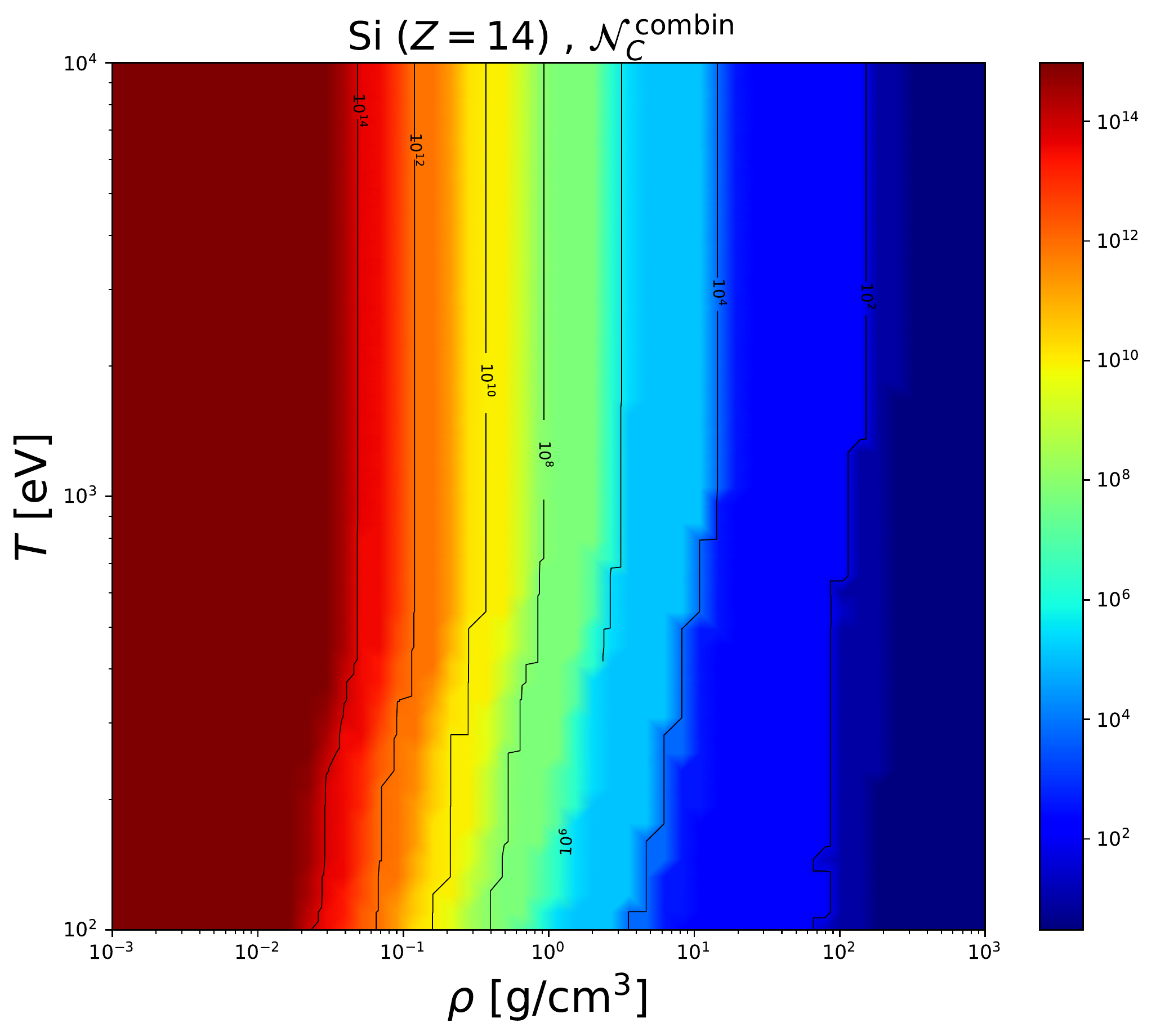}\includegraphics[scale=0.28]{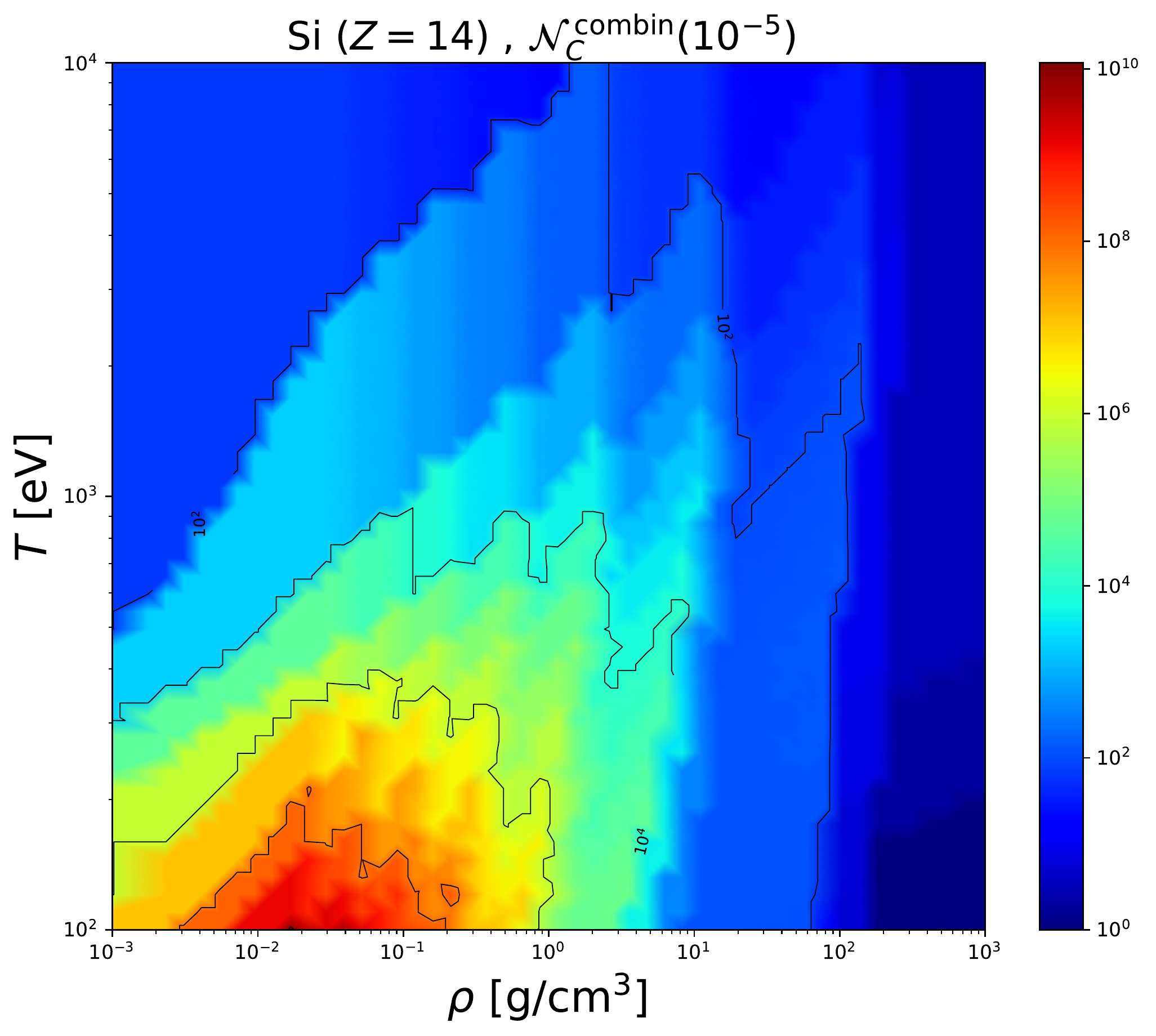}
\par\end{centering}
\begin{centering}
\includegraphics[scale=0.28]{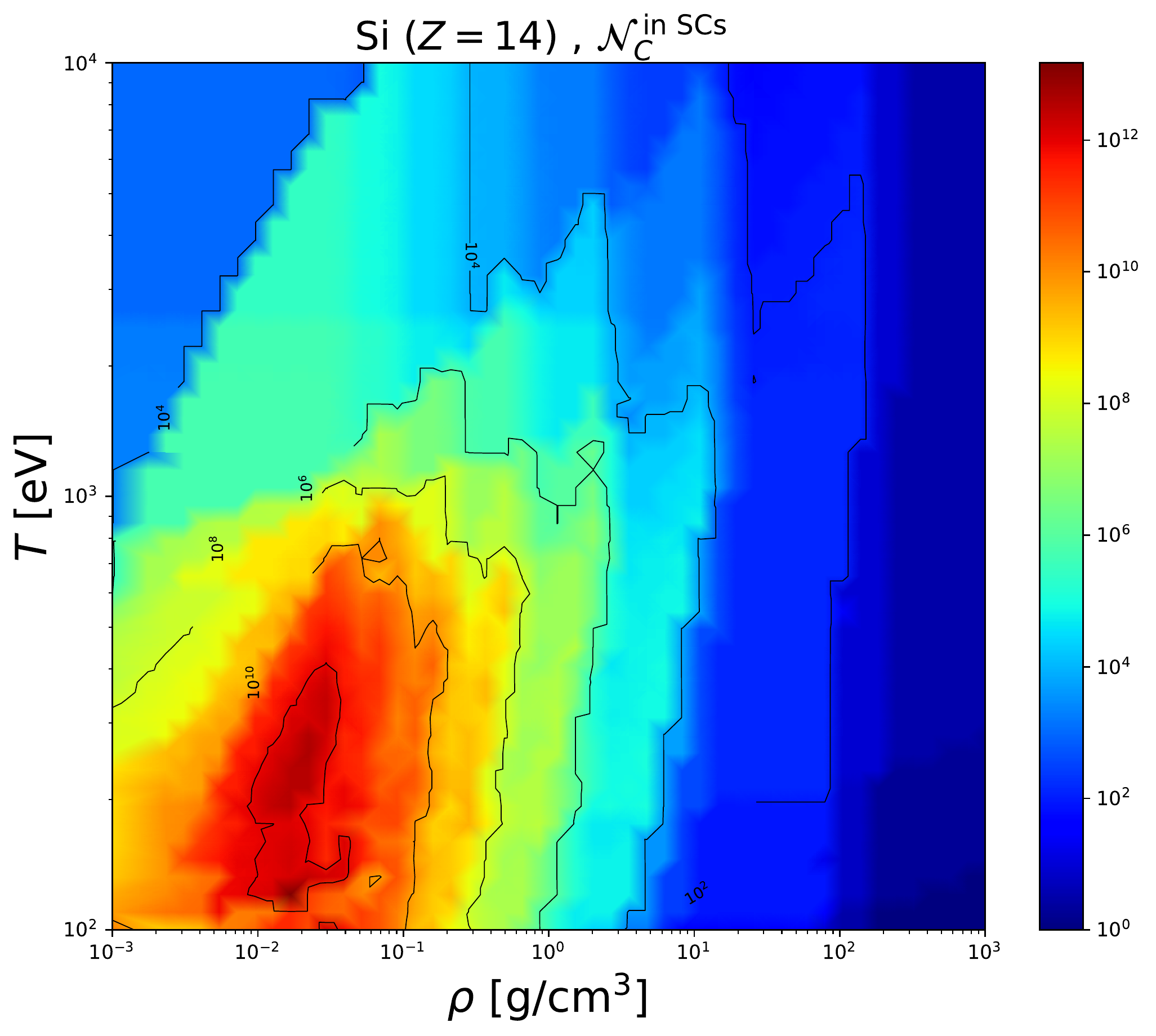}\includegraphics[scale=0.28]{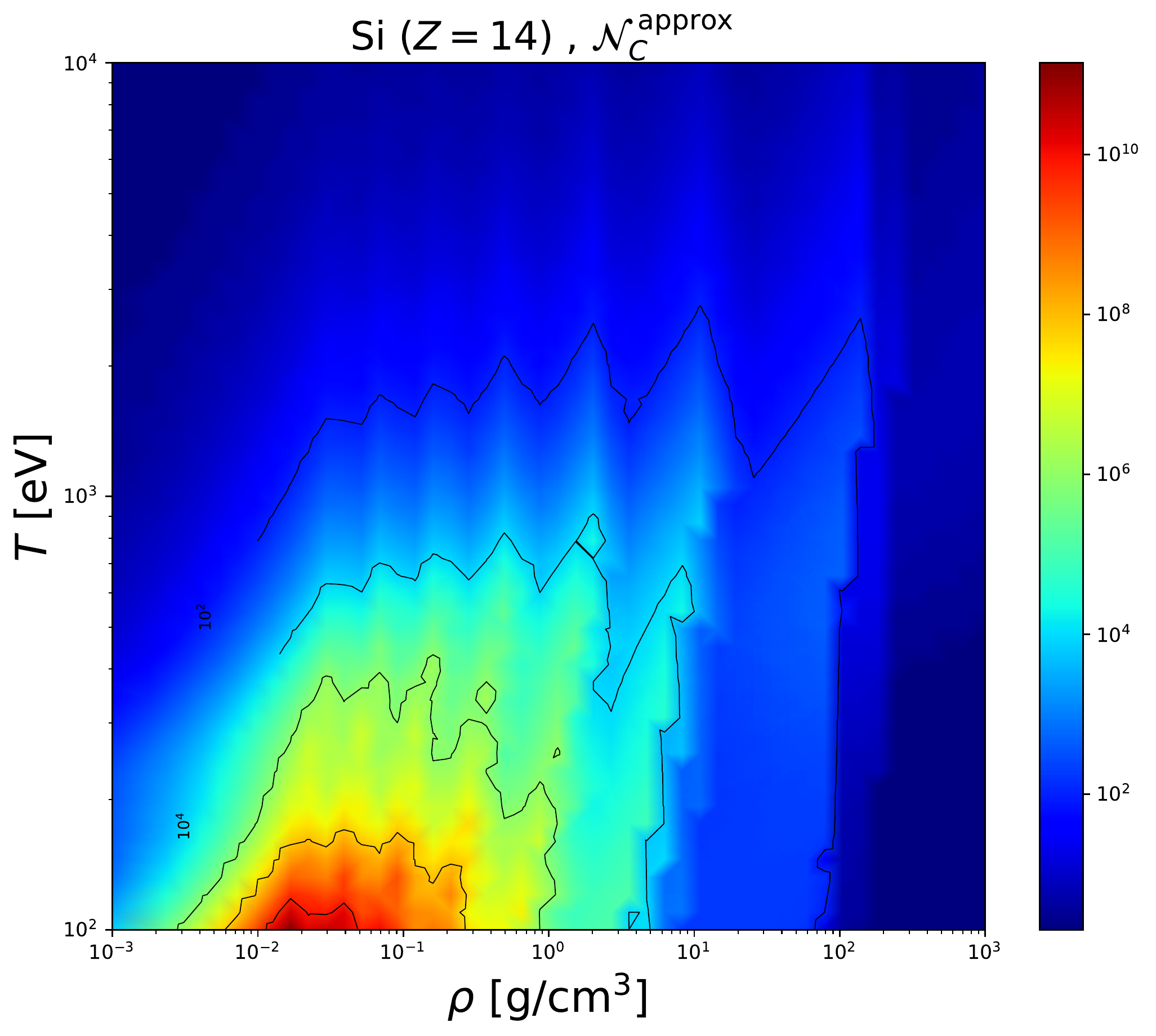}\includegraphics[scale=0.28]{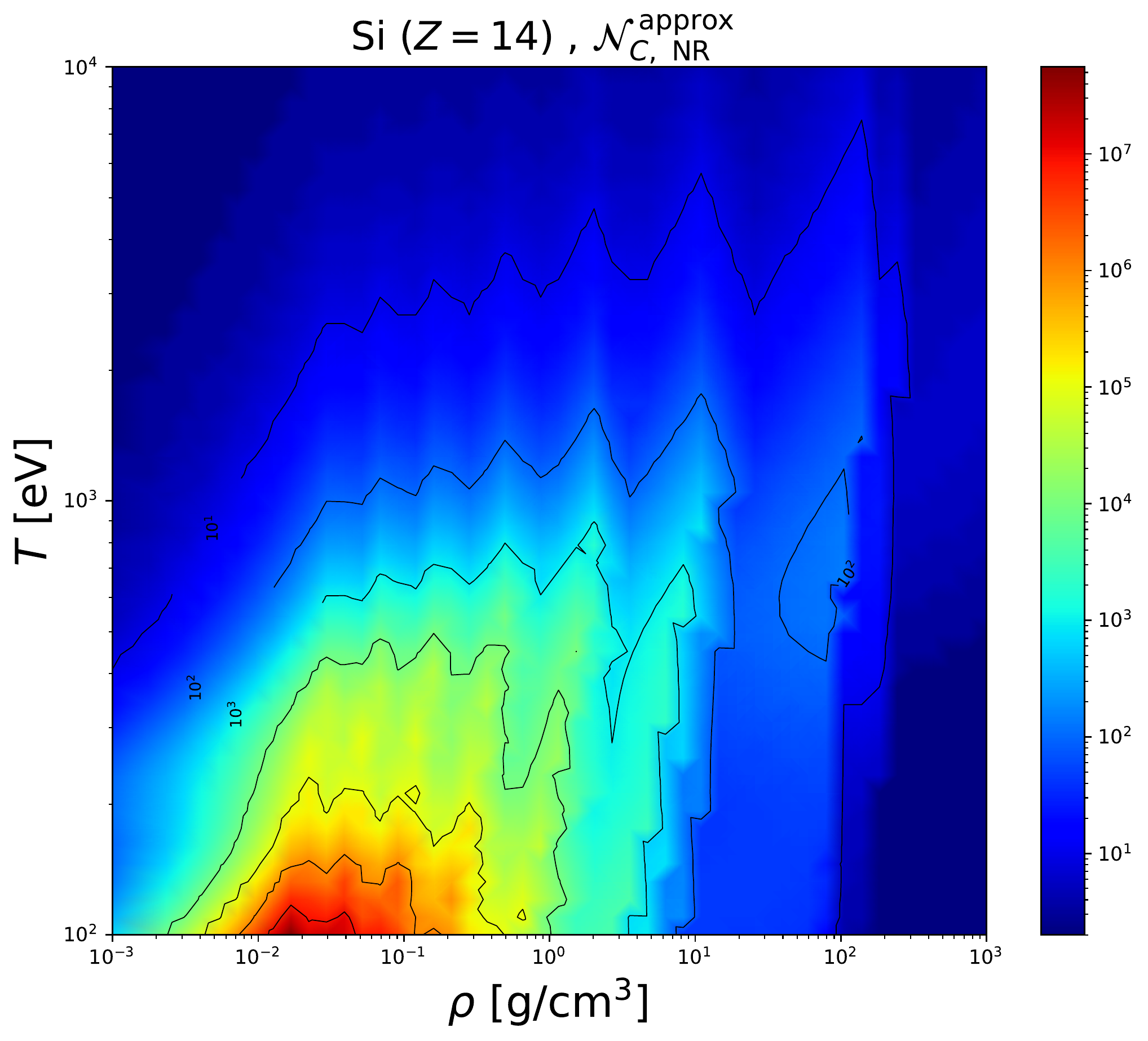} 
\par\end{centering}
\begin{centering}
\includegraphics[scale=0.28]{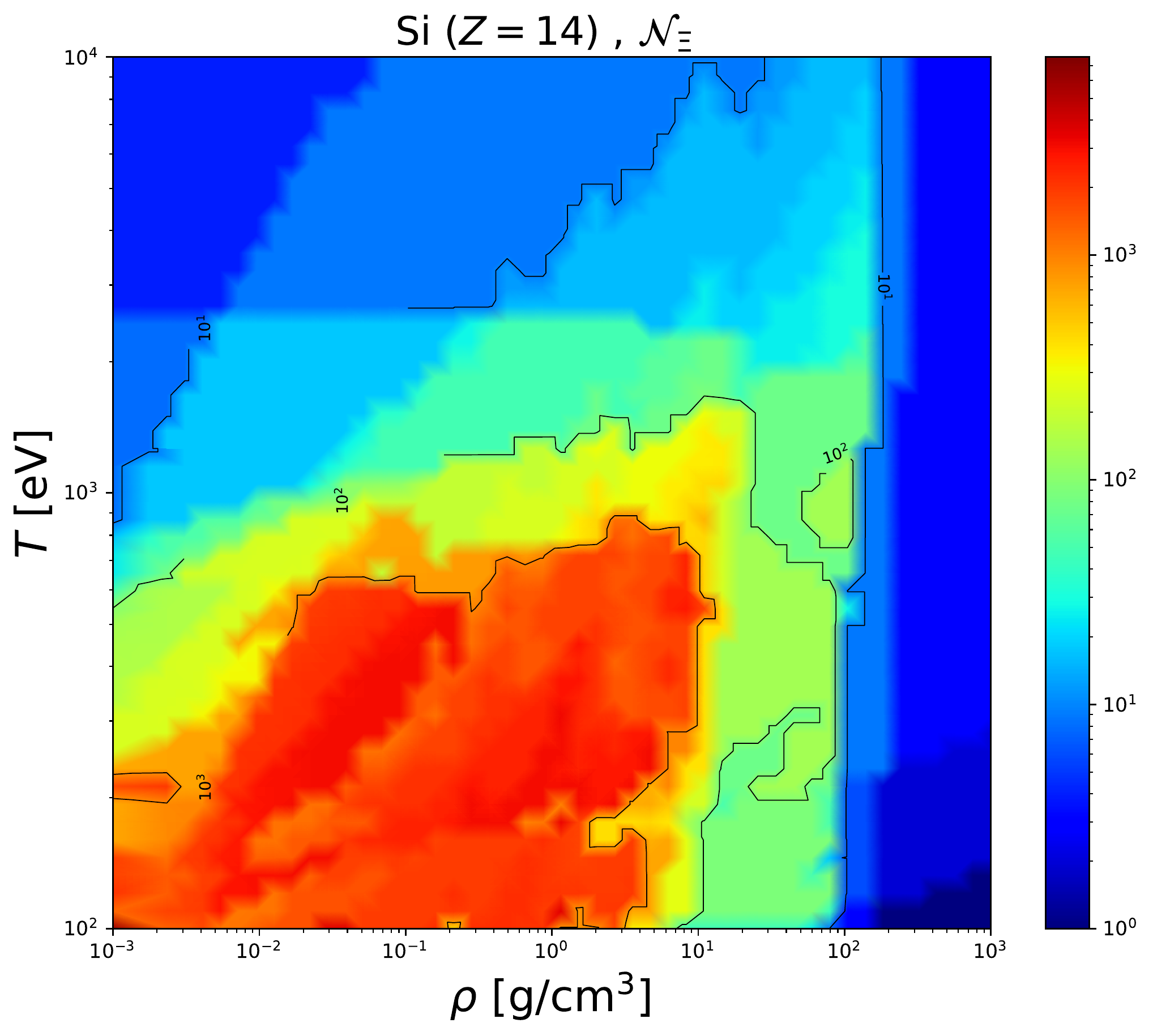}\includegraphics[scale=0.28]{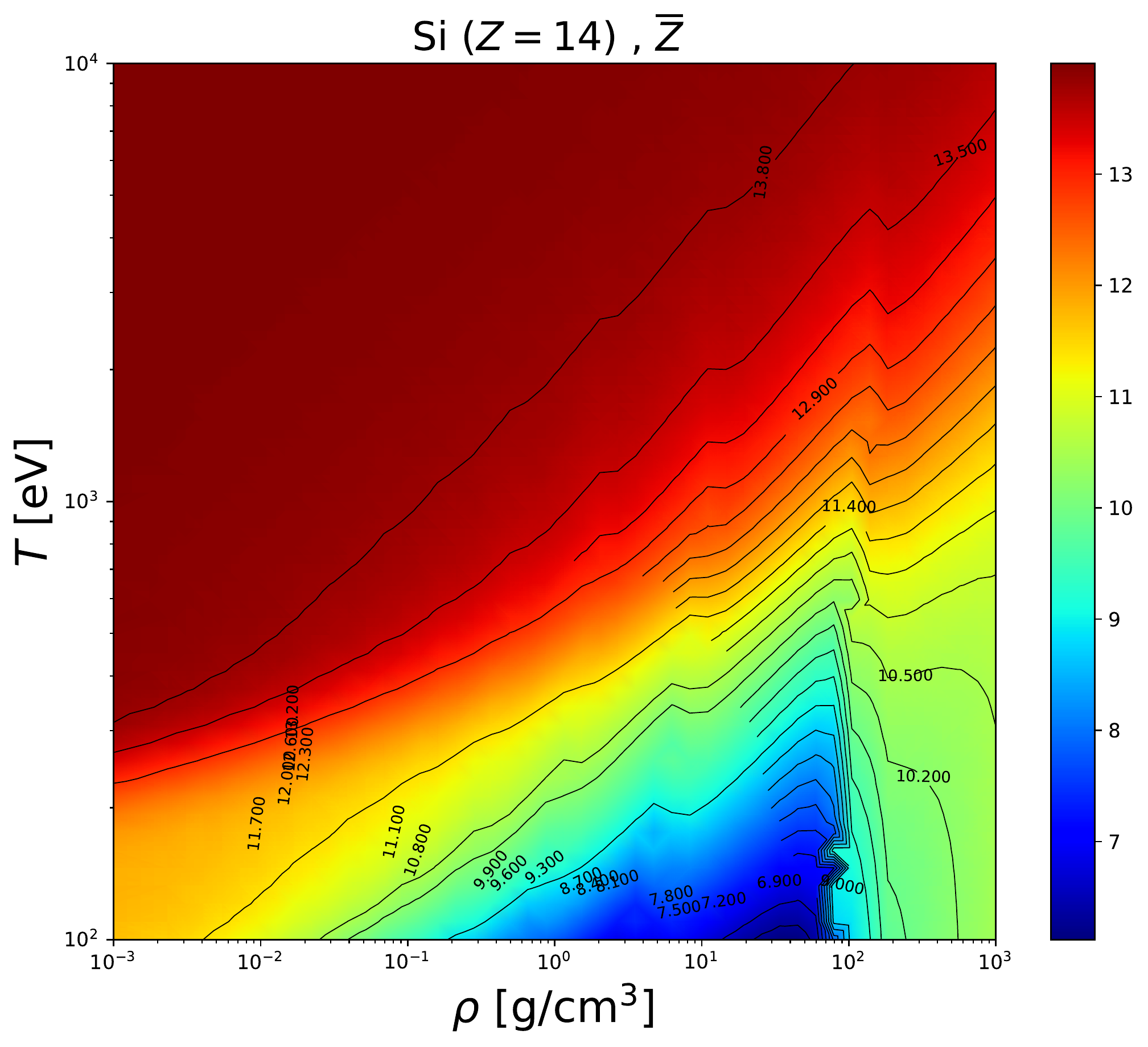}\includegraphics[scale=0.28]{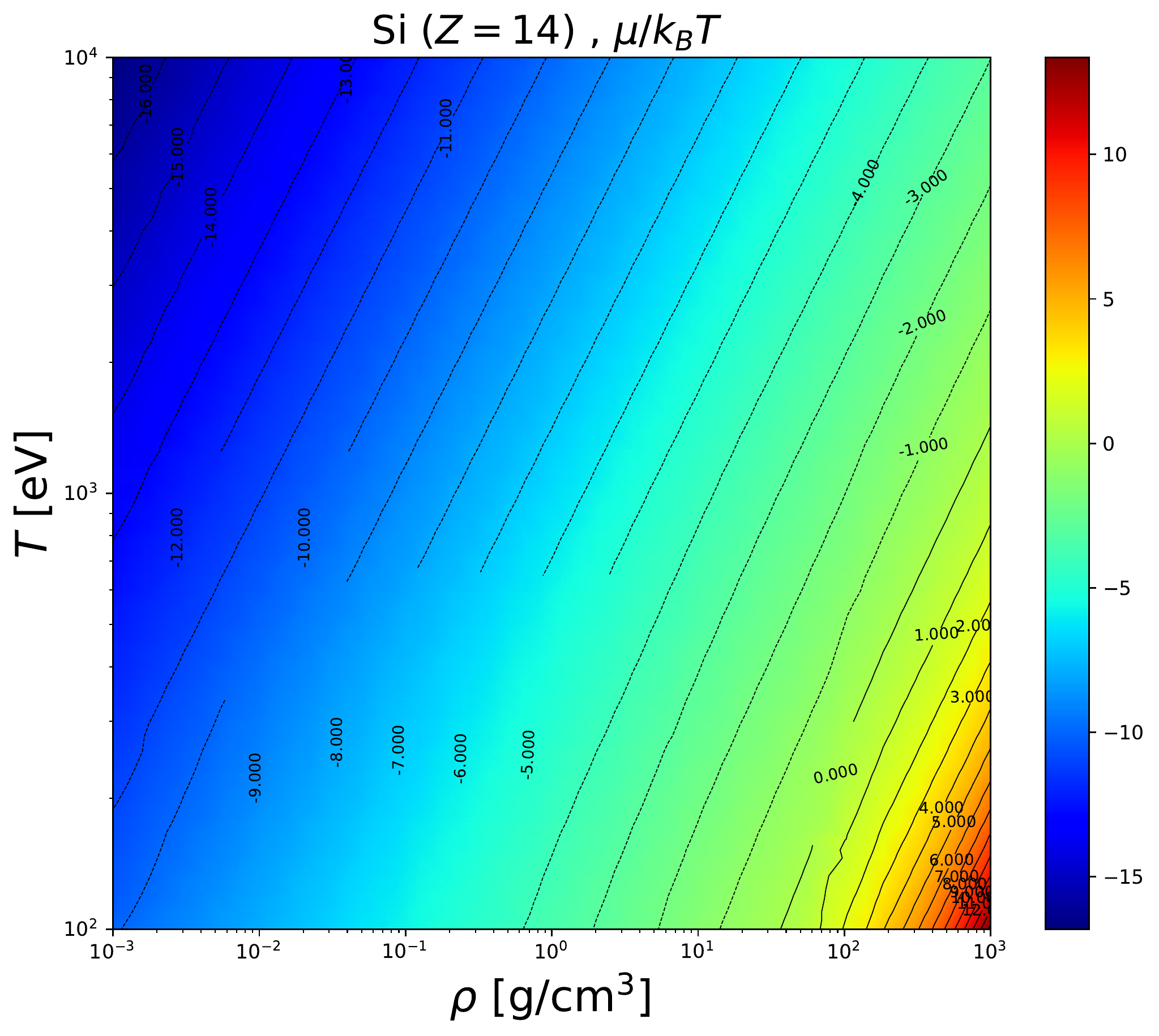} 
\par\end{centering}
\caption{(Color online) Various color plots for Silicone (Z=14), as a function of temperature
and density (left to right, top to bottom): the number of bound shells
$N_{\text{shells}}$, the combinatoric number of relativistic configurations
$\mathcal{N}_{C}^{\text{combin}}$ over all ionization levels (eq.
\ref{eq:NC_COMBIN}) and over all ionization levels with probability
larger than $10^{-5}$ (eq. \ref{eq:NC_COMBIN_P}), the number of
superconfigurations in a converged STA calculation $\mathcal{N}_{\Xi}$,
the number of configurations within superconfigurations $\mathcal{N}_{C}^{\text{in SCs}}$
(eq. \ref{eq:nc_insc} with $p=10^{-7}$), approximated number of
populated relativistic $\mathcal{N}_{C}^{\text{approx}}$ and non-relativistic
$\mathcal{N}_{C,\text{NR}}^{\text{approx}}$ configurations, the average
ionization $\overline{Z}$ and the normalized chemical potential $\mu/k_{B}T$.\label{fig:si_meshes}}
\end{figure*}

\begin{figure*}
\begin{centering}
\includegraphics[scale=0.28]{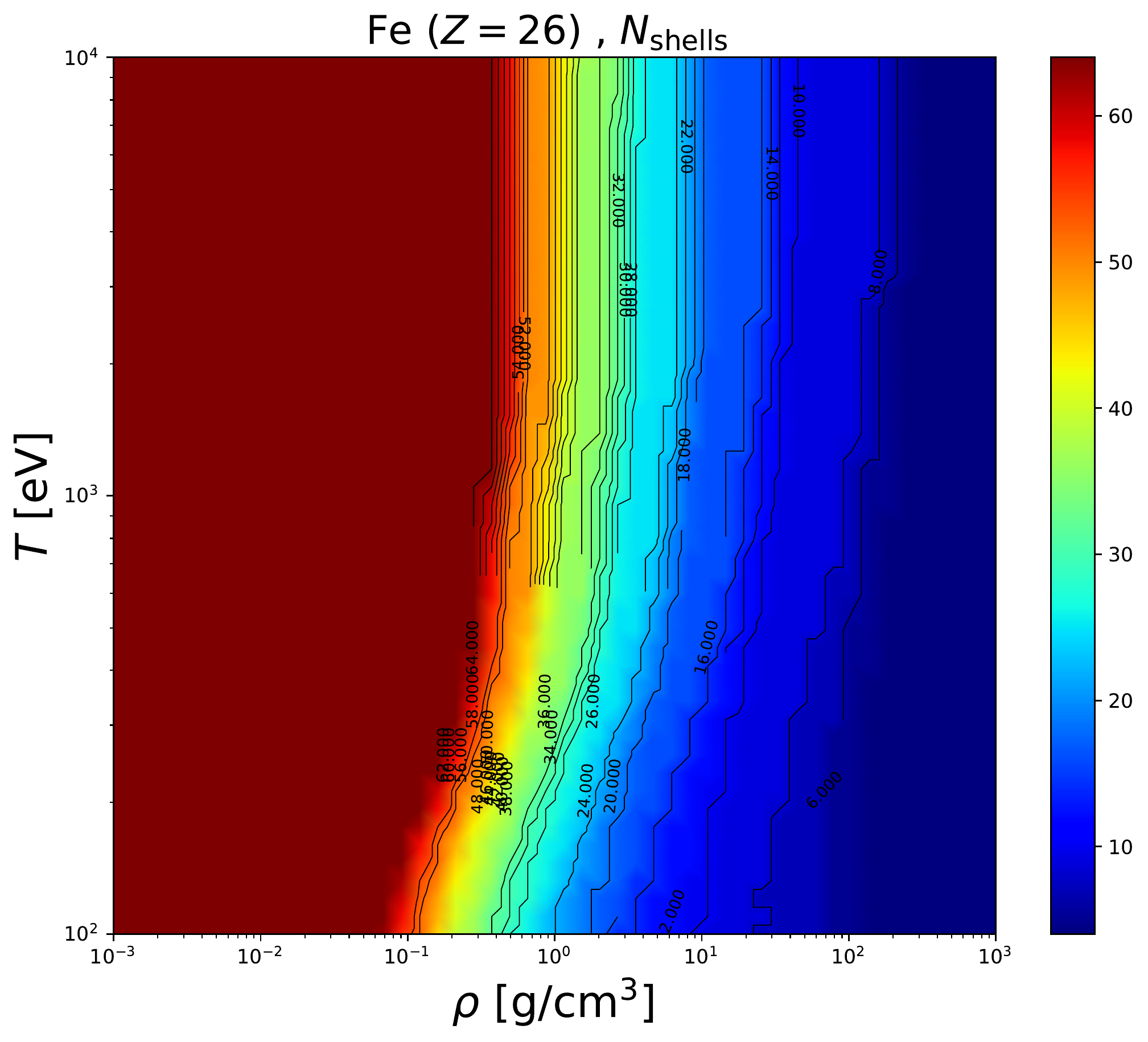}\includegraphics[scale=0.28]{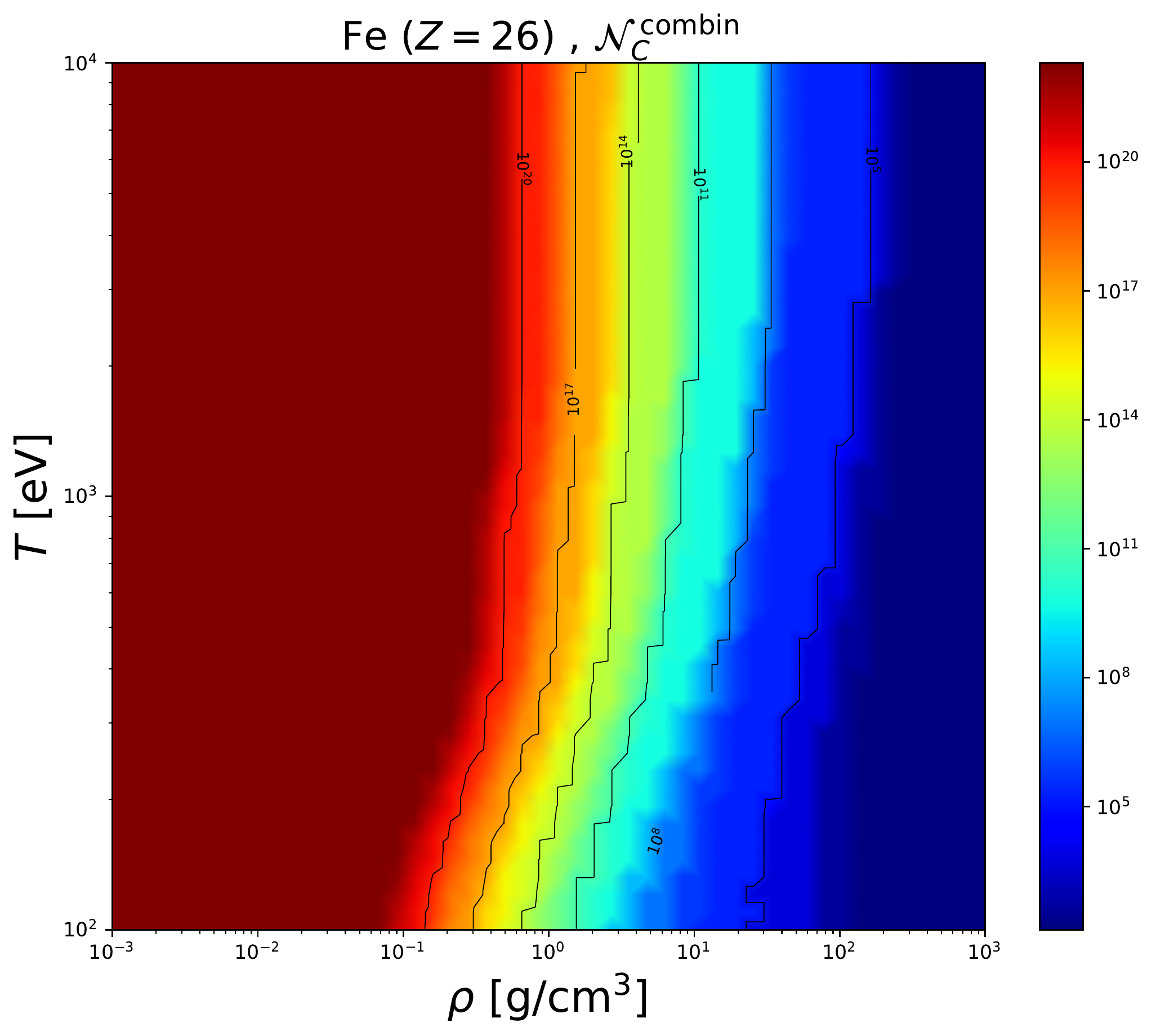}\includegraphics[scale=0.28]{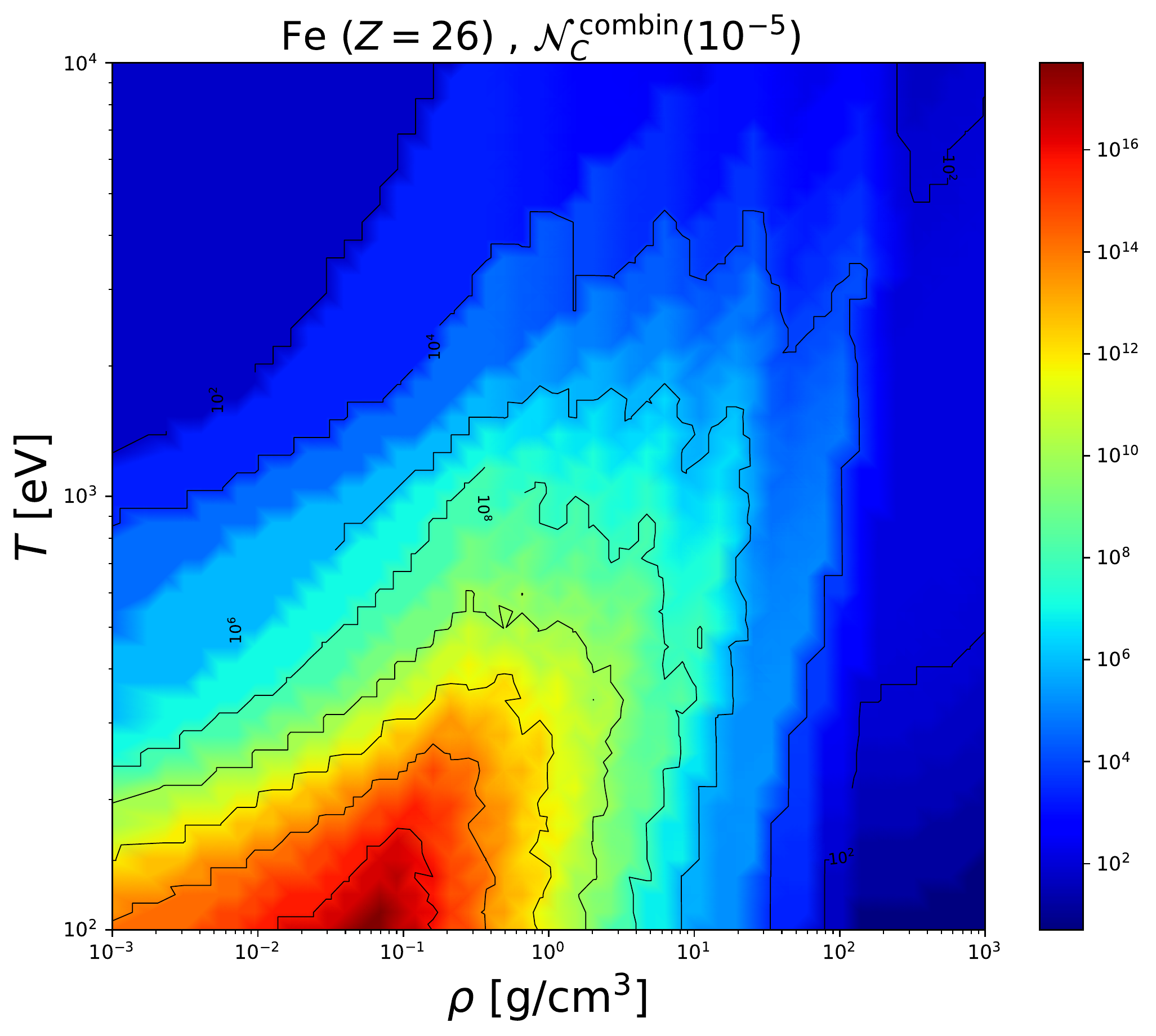}
\par\end{centering}
\begin{centering}
\includegraphics[scale=0.28]{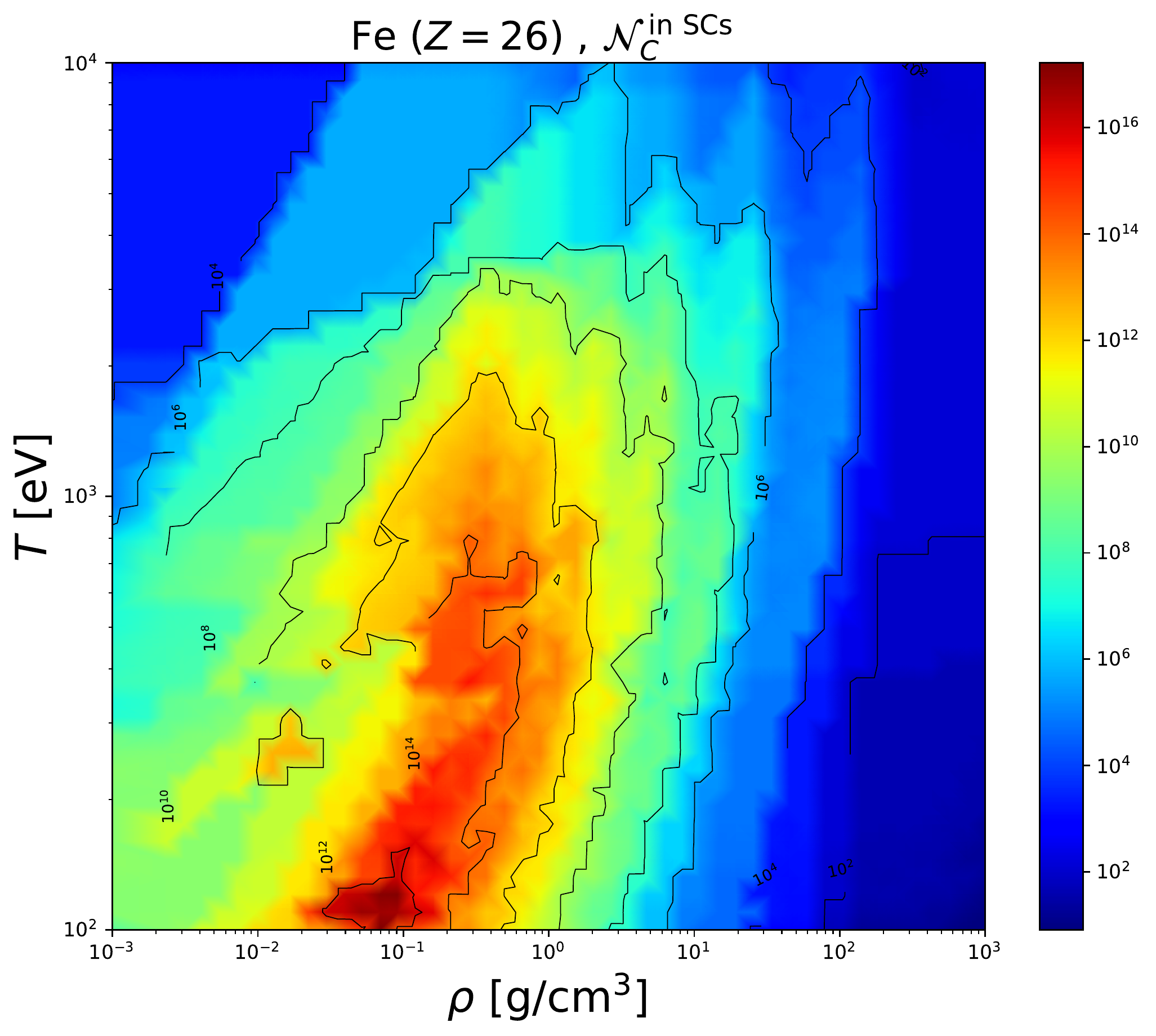}\includegraphics[scale=0.28]{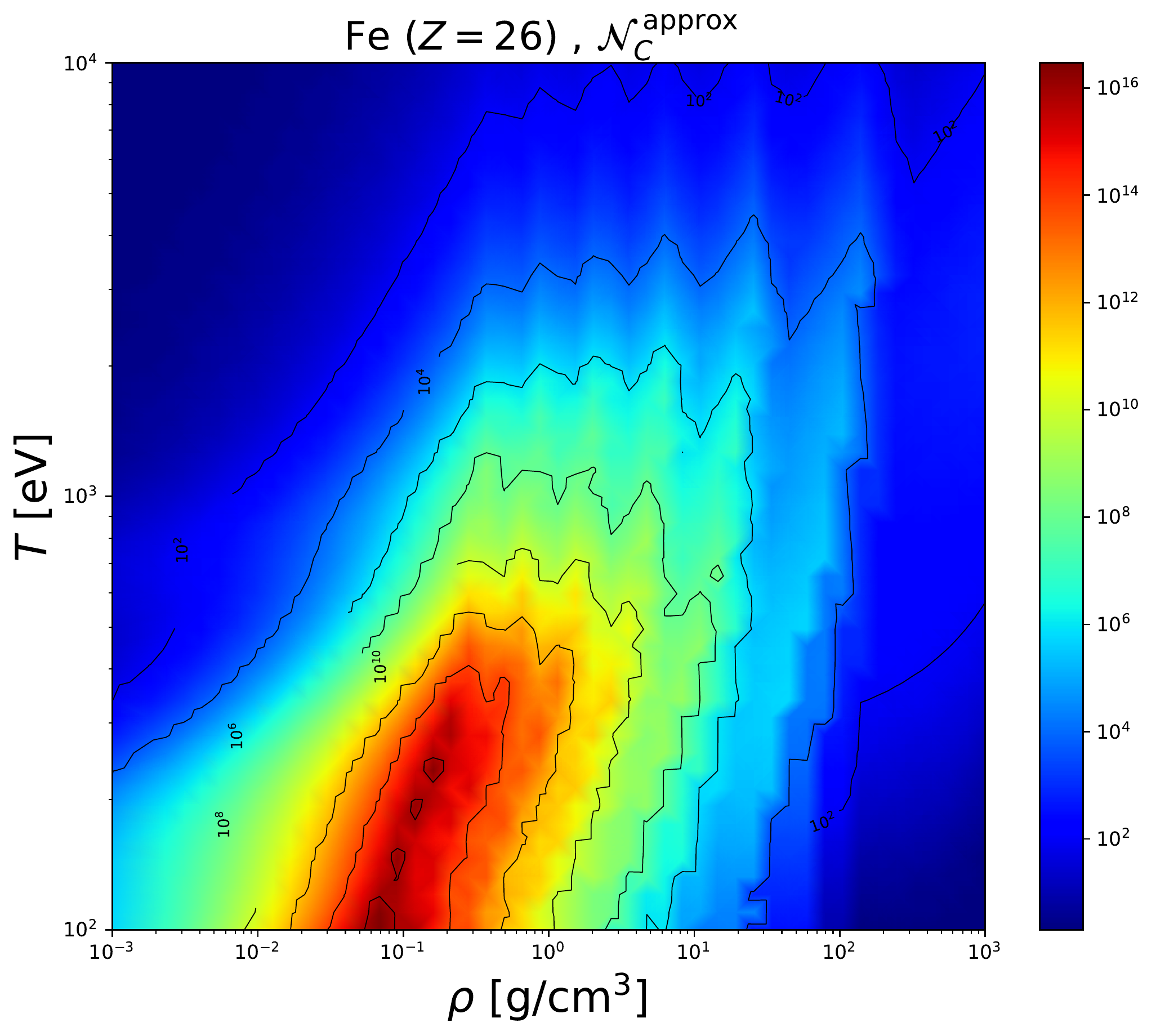}\includegraphics[scale=0.28]{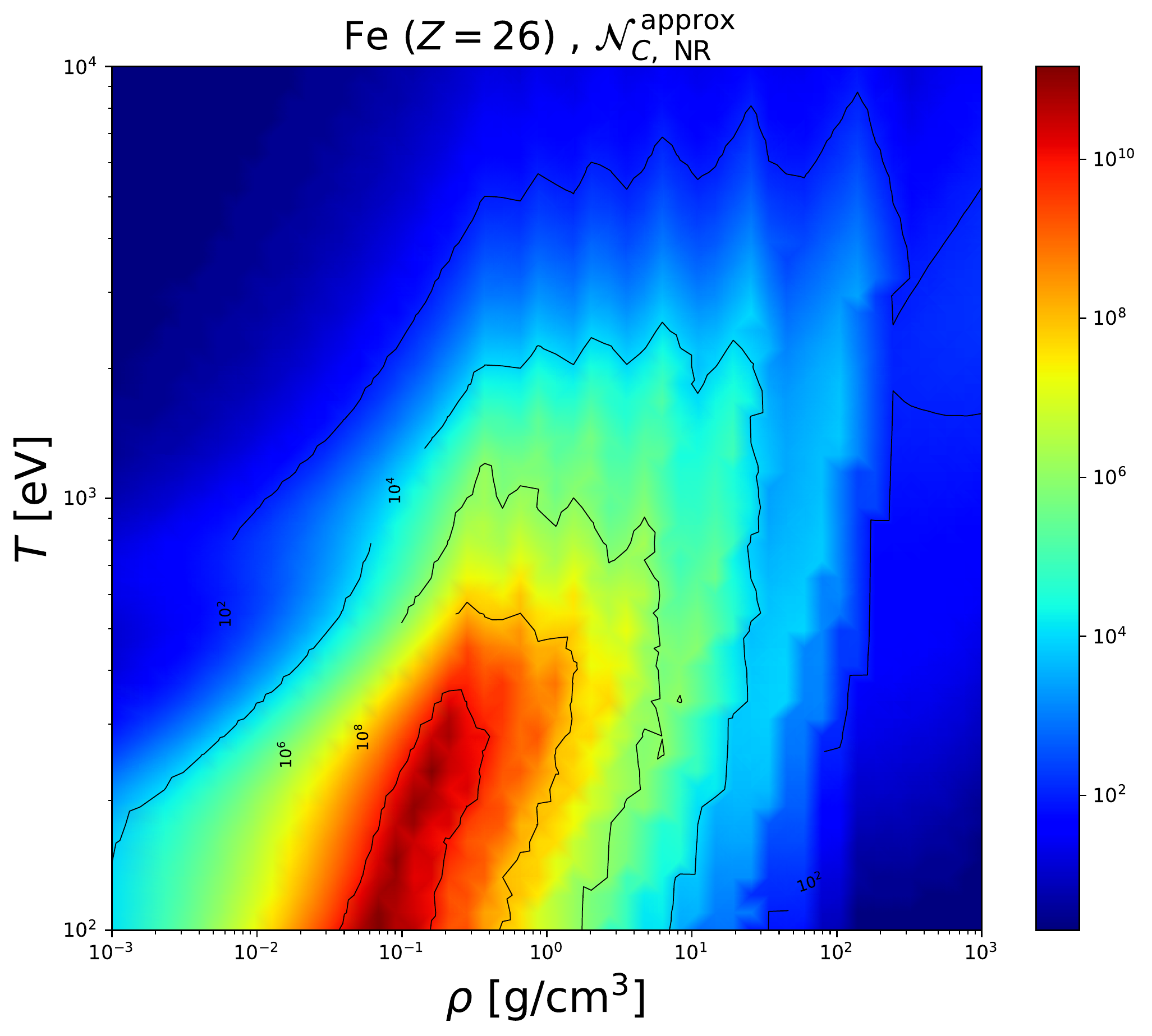} 
\par\end{centering}
\begin{centering}
\includegraphics[scale=0.28]{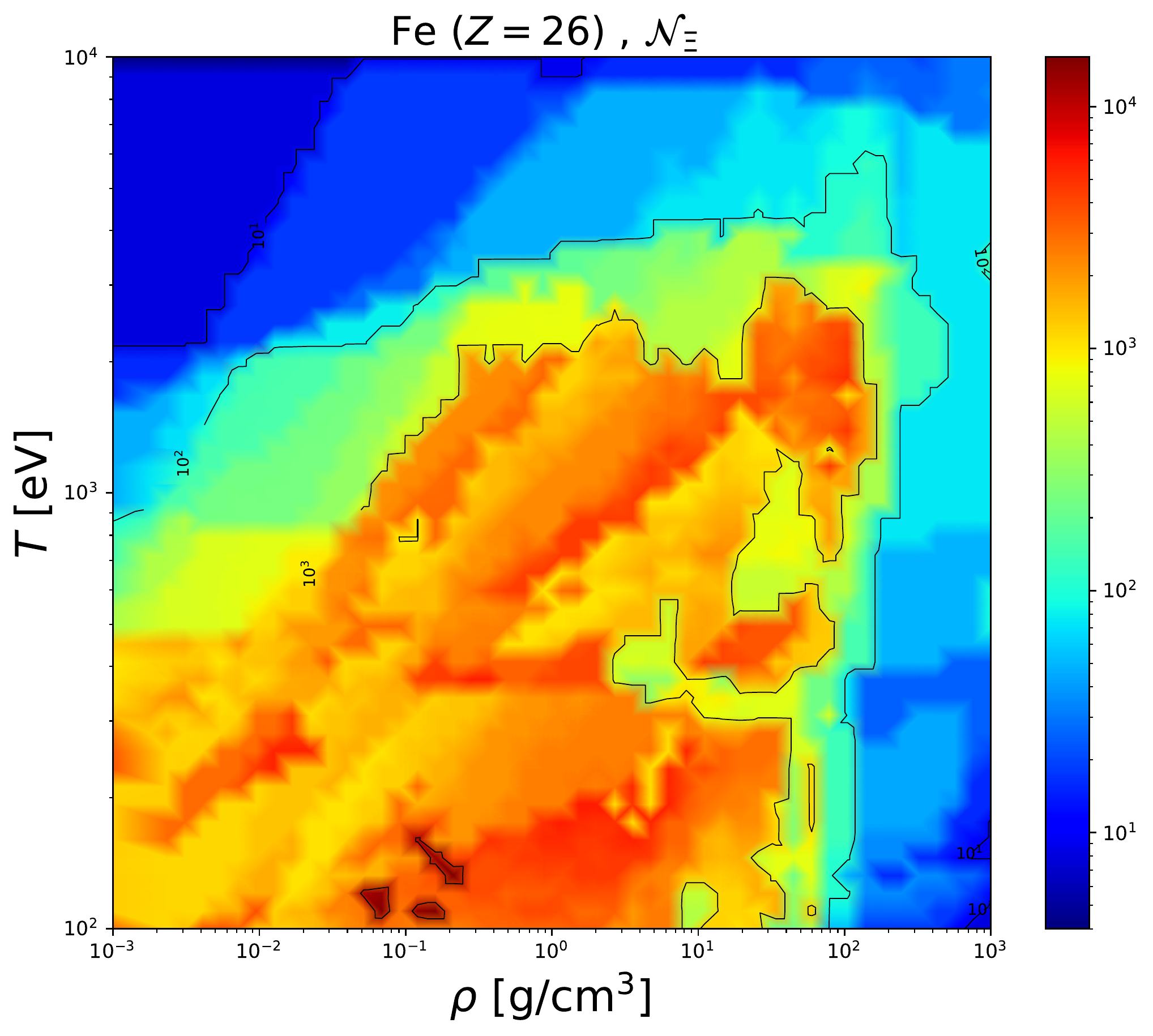}\includegraphics[scale=0.28]{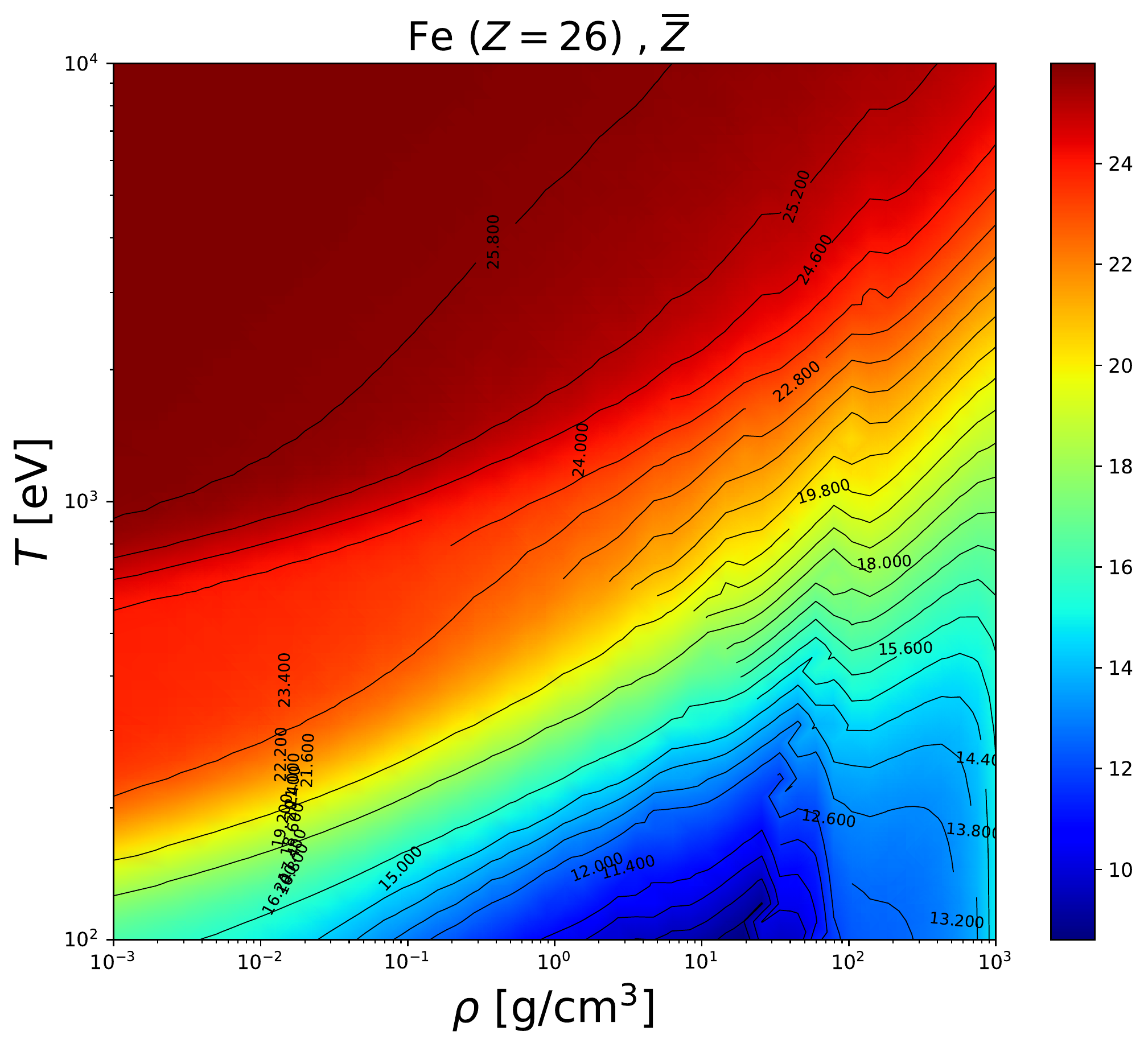}\includegraphics[scale=0.28]{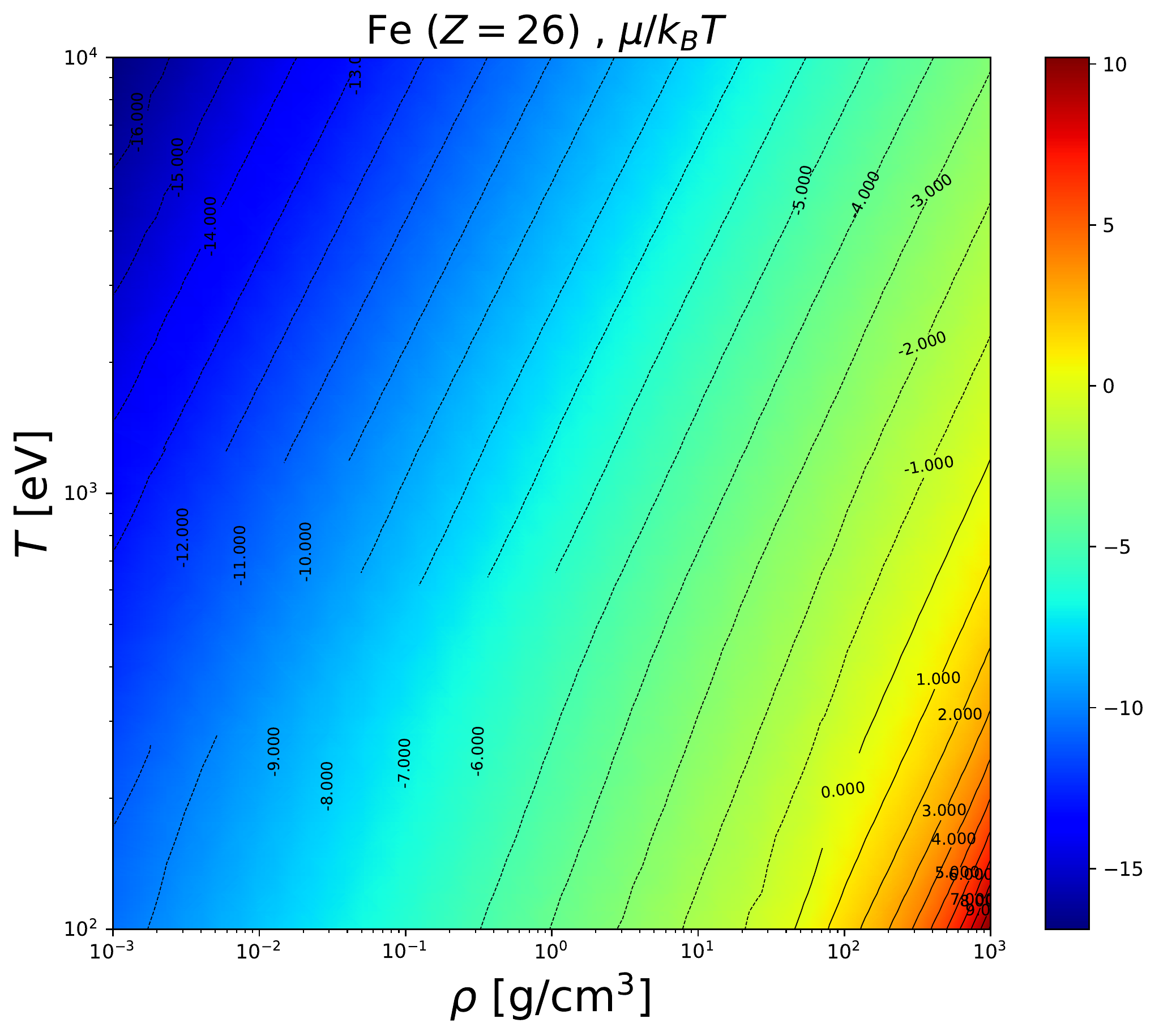} 
\par\end{centering}
\caption{(Color online) Same as Fig. \ref{fig:si_meshes}, for Iron (Z=26).\label{fig:fe_meshes}}
\end{figure*}

\begin{figure*}
\begin{centering}
\includegraphics[scale=0.28]{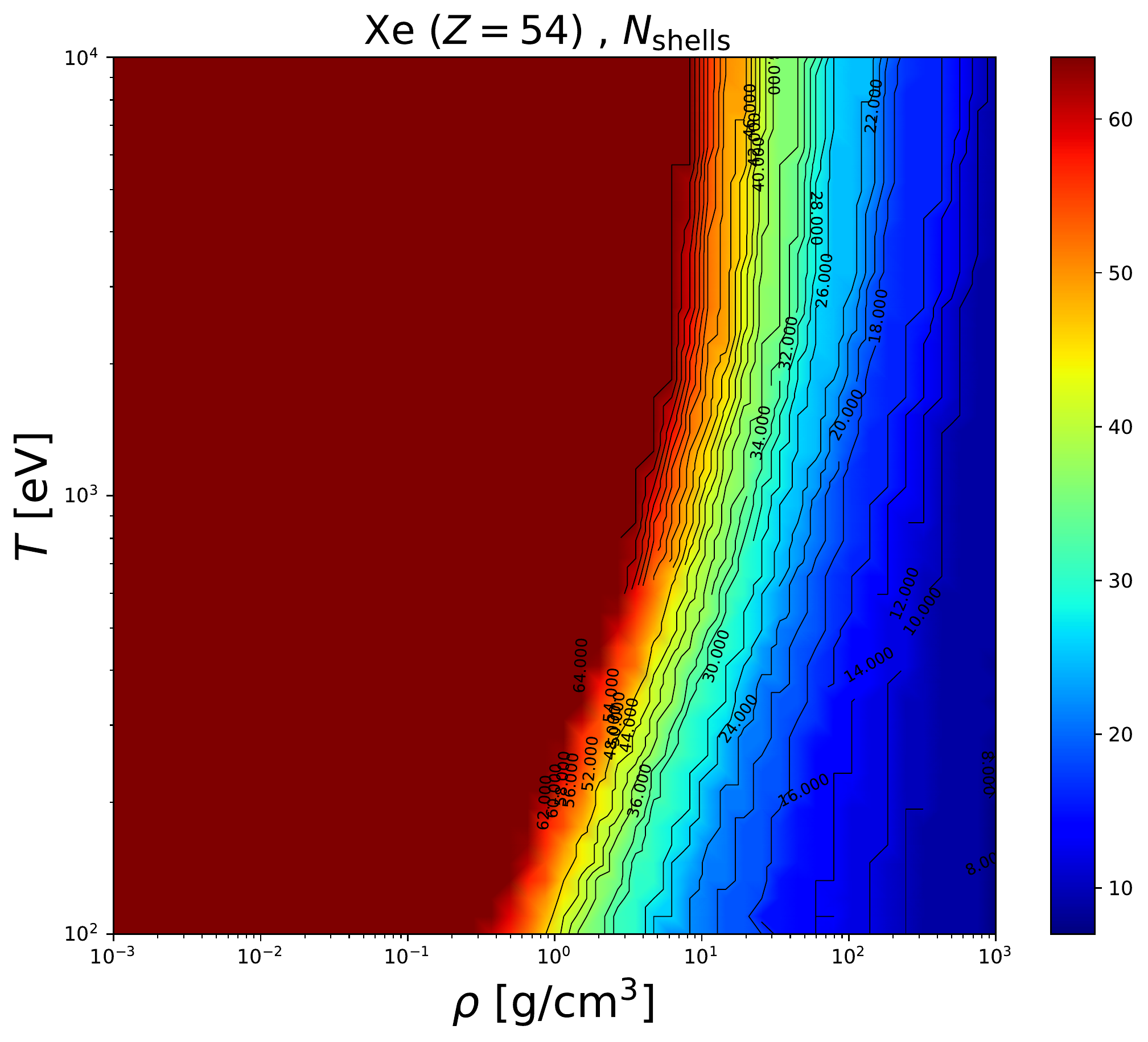}\includegraphics[scale=0.28]{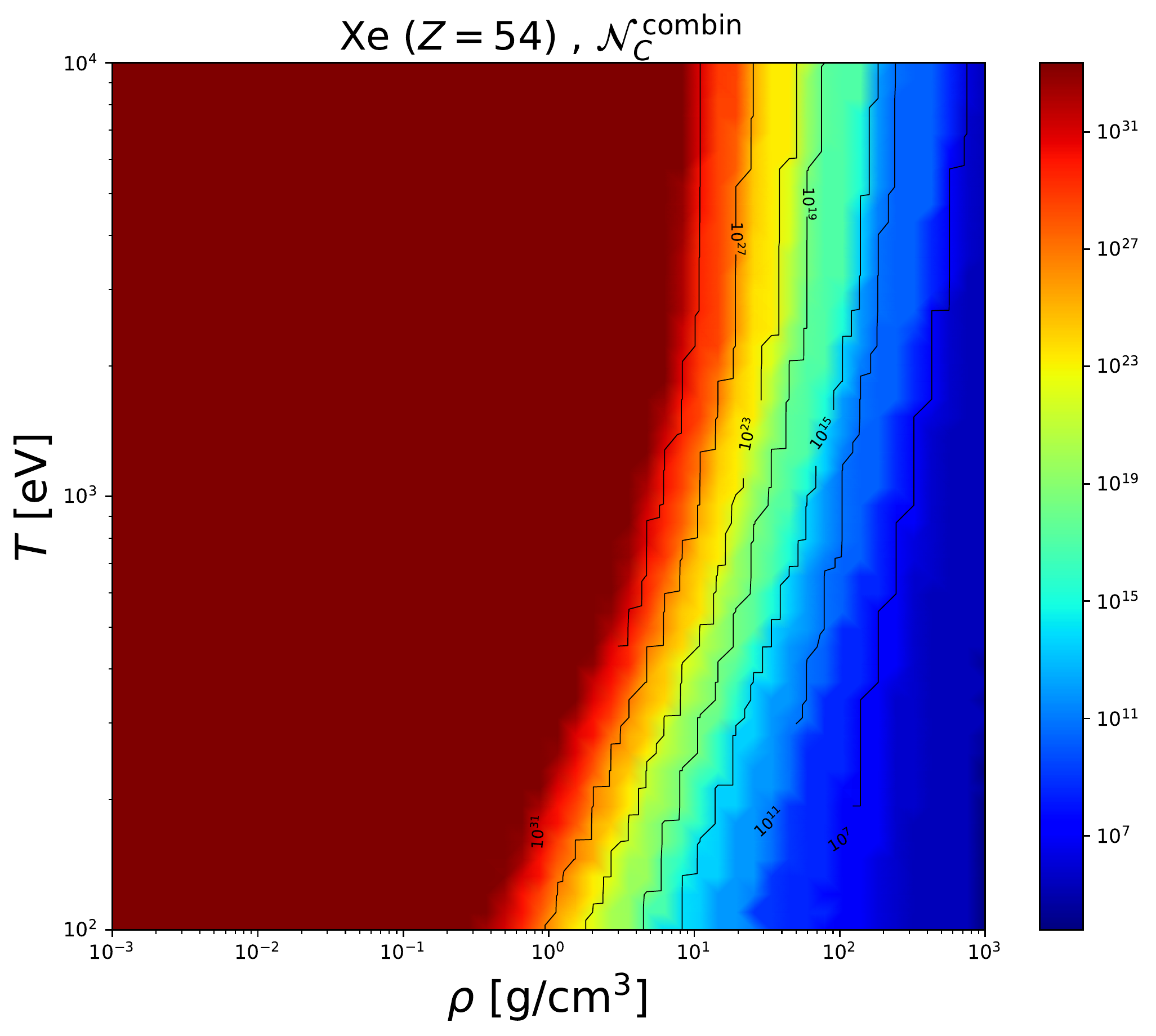}\includegraphics[scale=0.28]{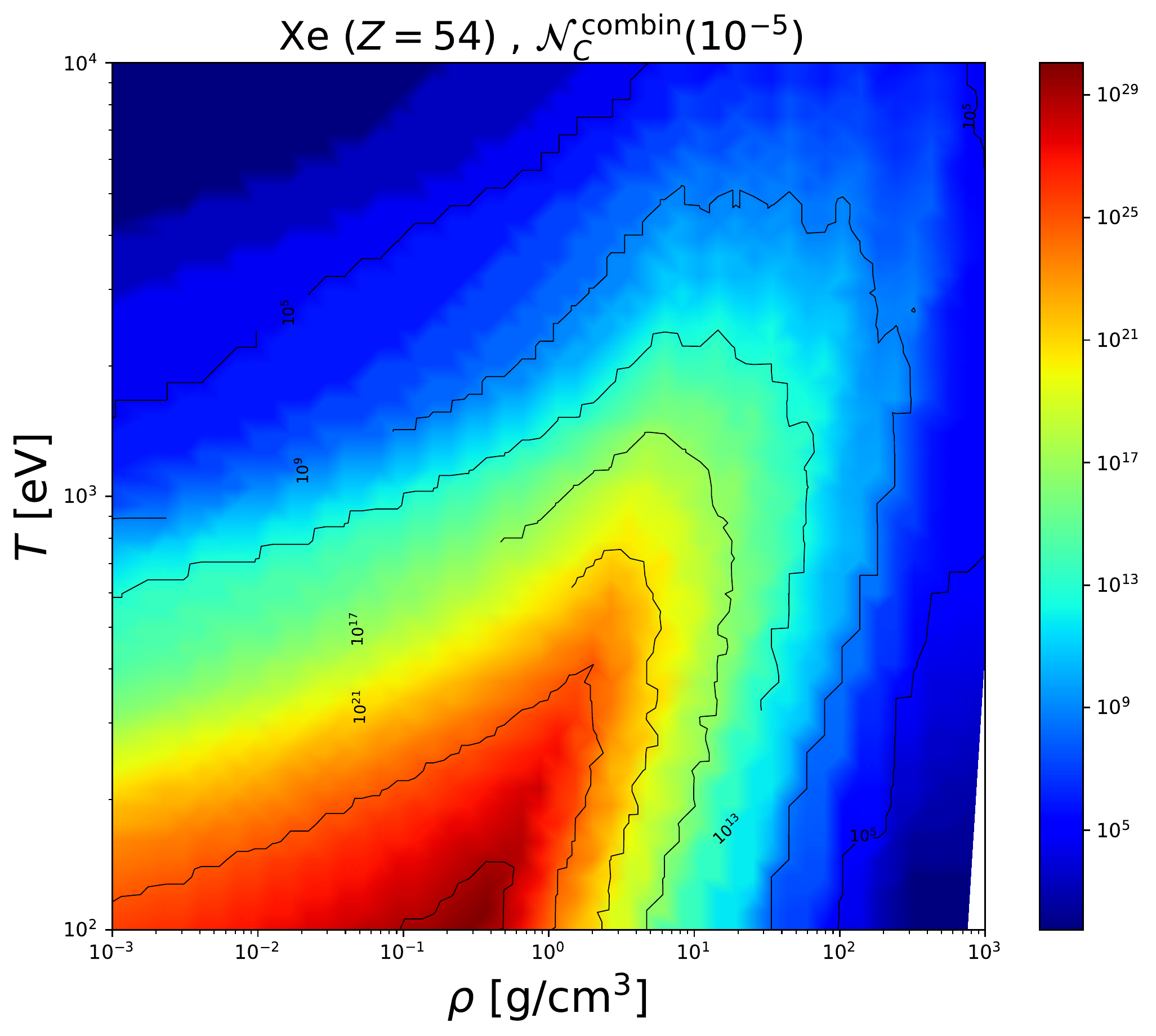}
\par\end{centering}
\begin{centering}
\includegraphics[scale=0.28]{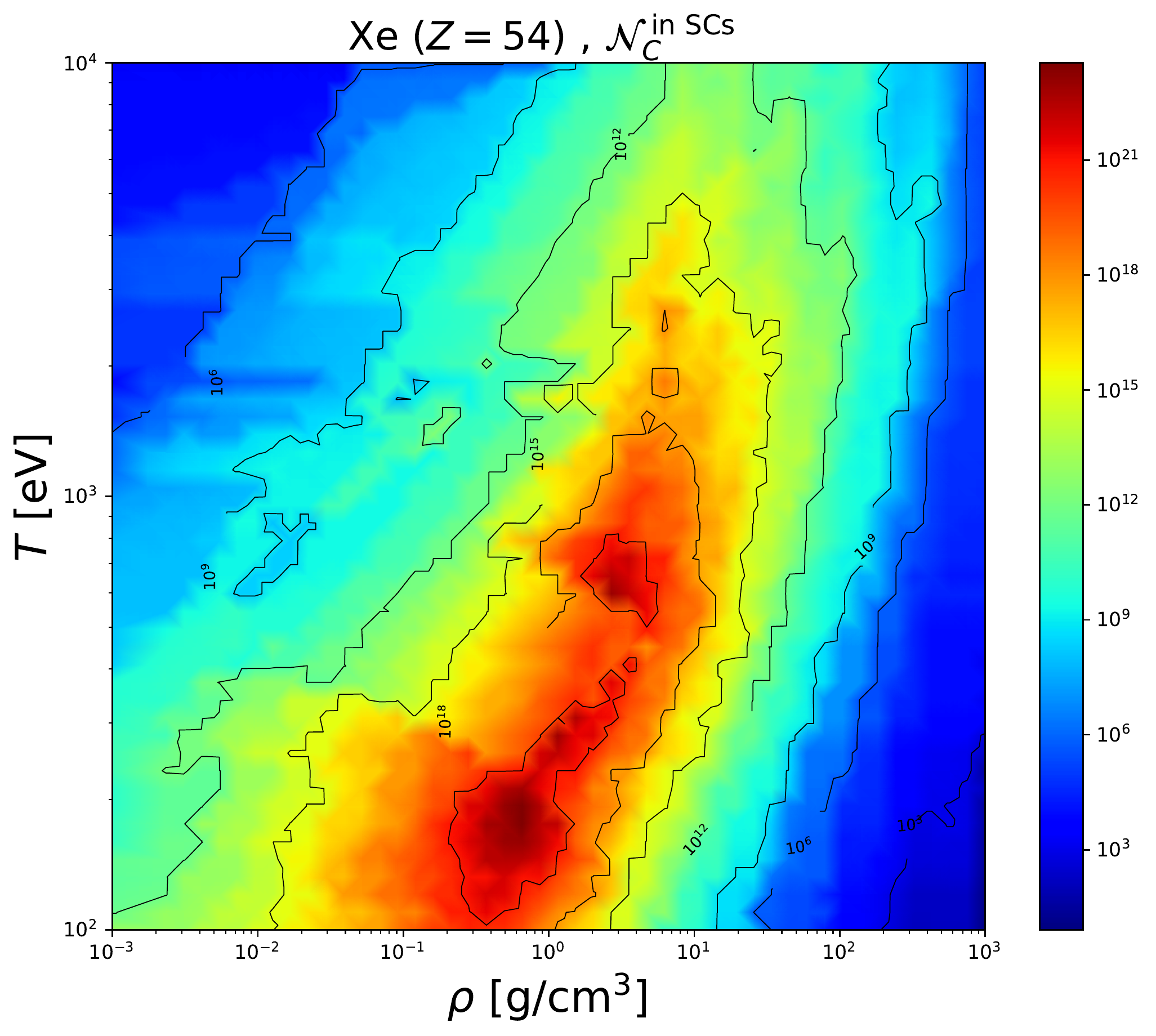}\includegraphics[scale=0.28]{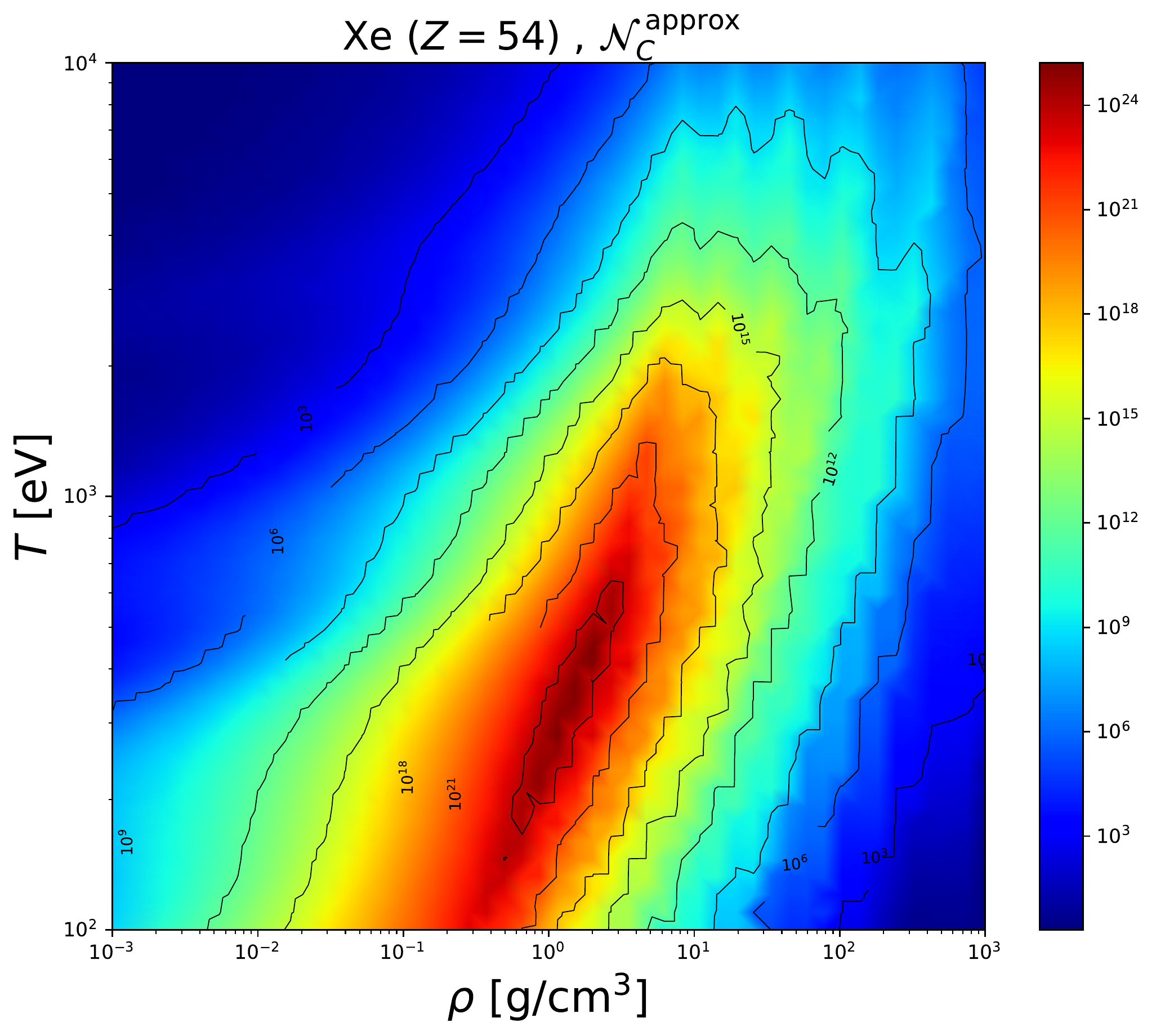}\includegraphics[scale=0.28]{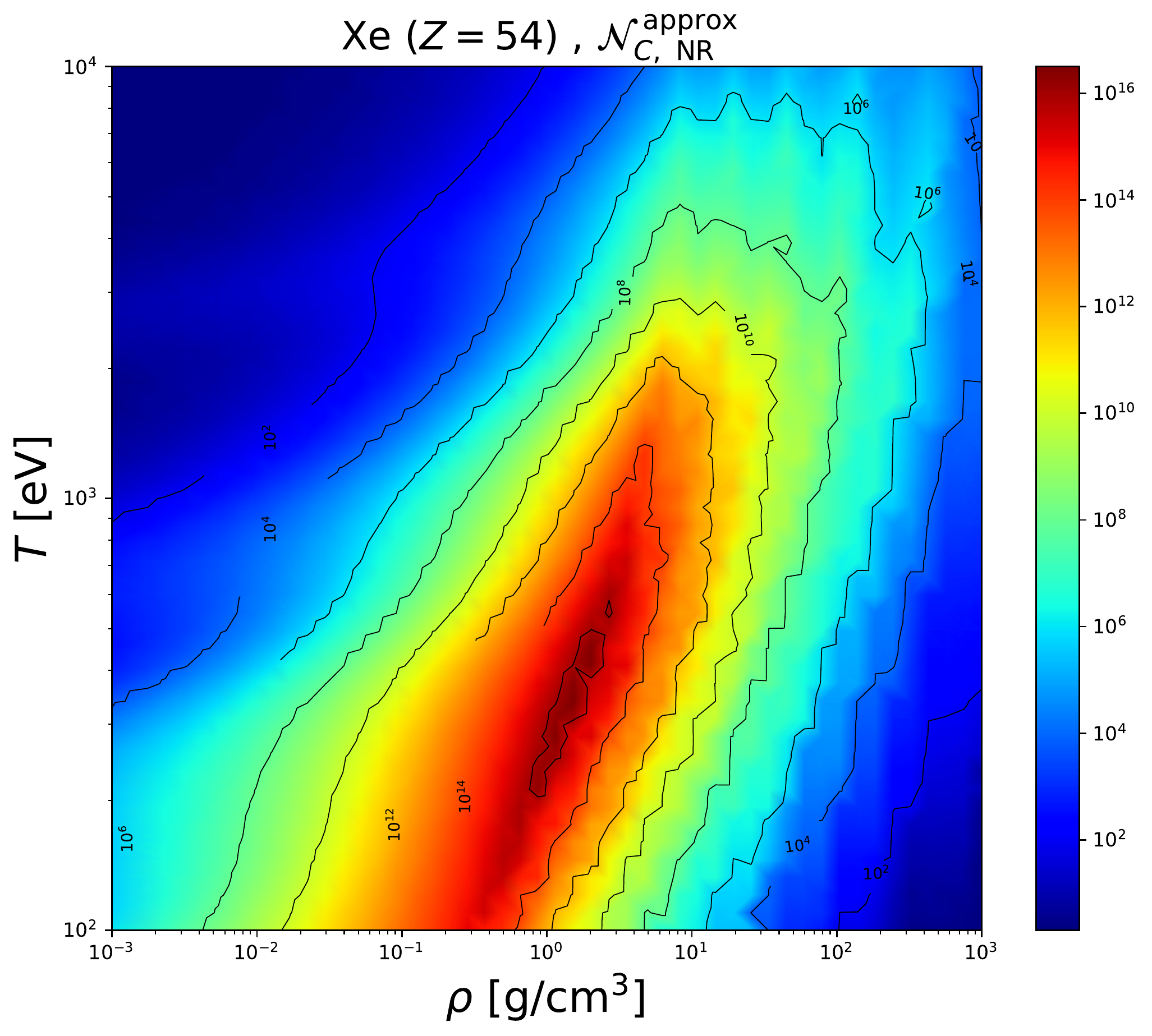} 
\par\end{centering}
\begin{centering}
\includegraphics[scale=0.28]{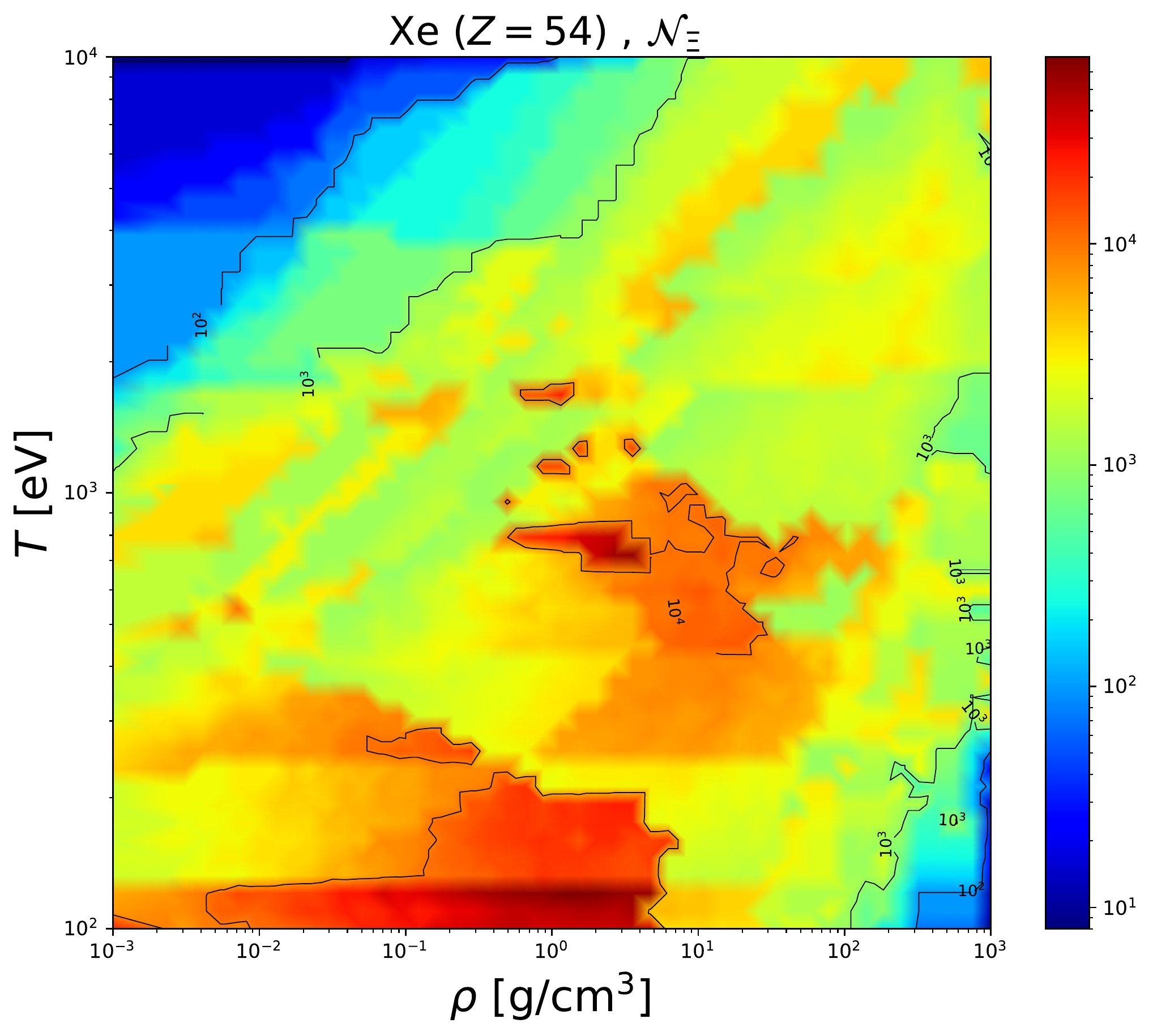}\includegraphics[scale=0.28]{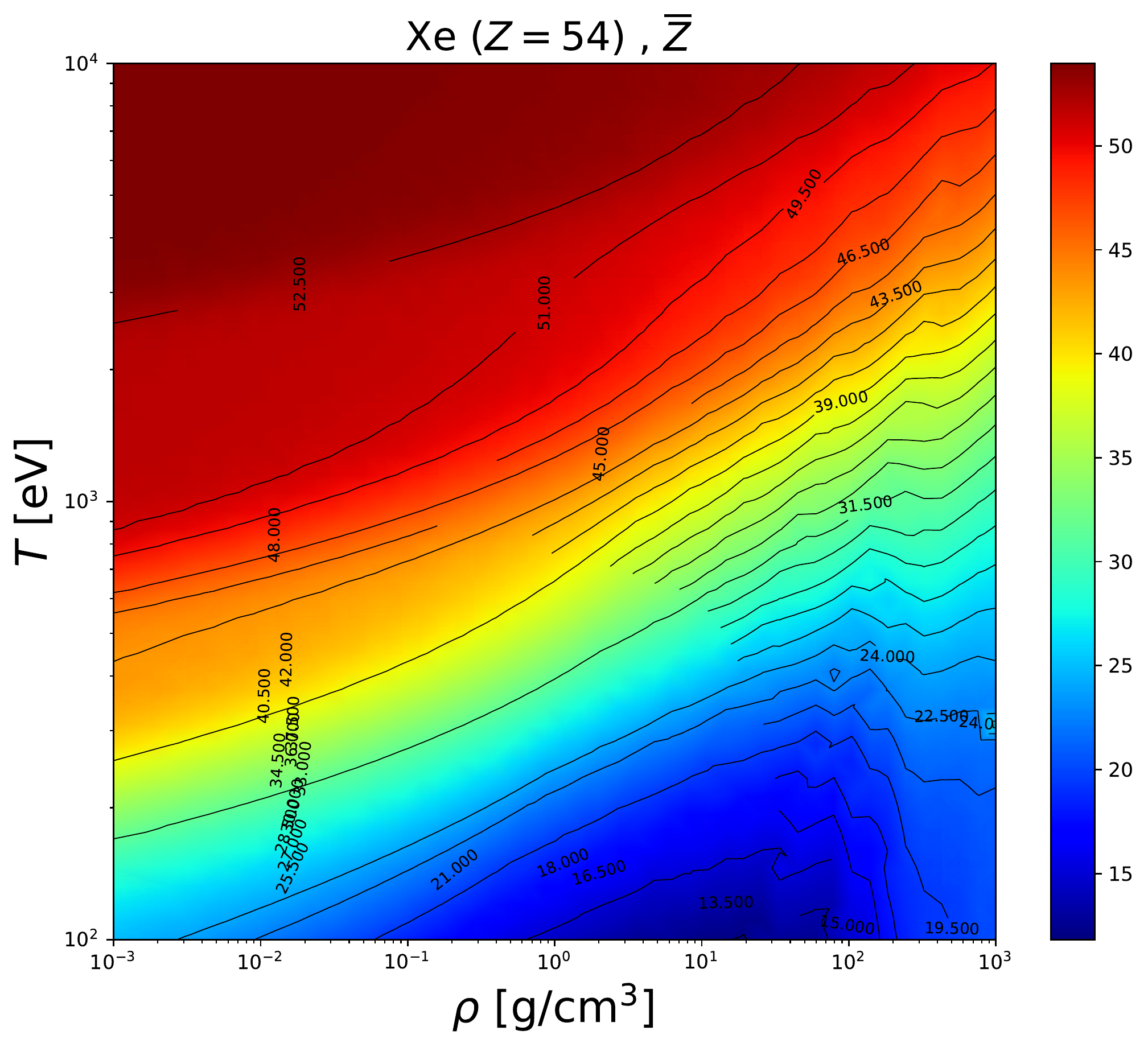}\includegraphics[scale=0.28]{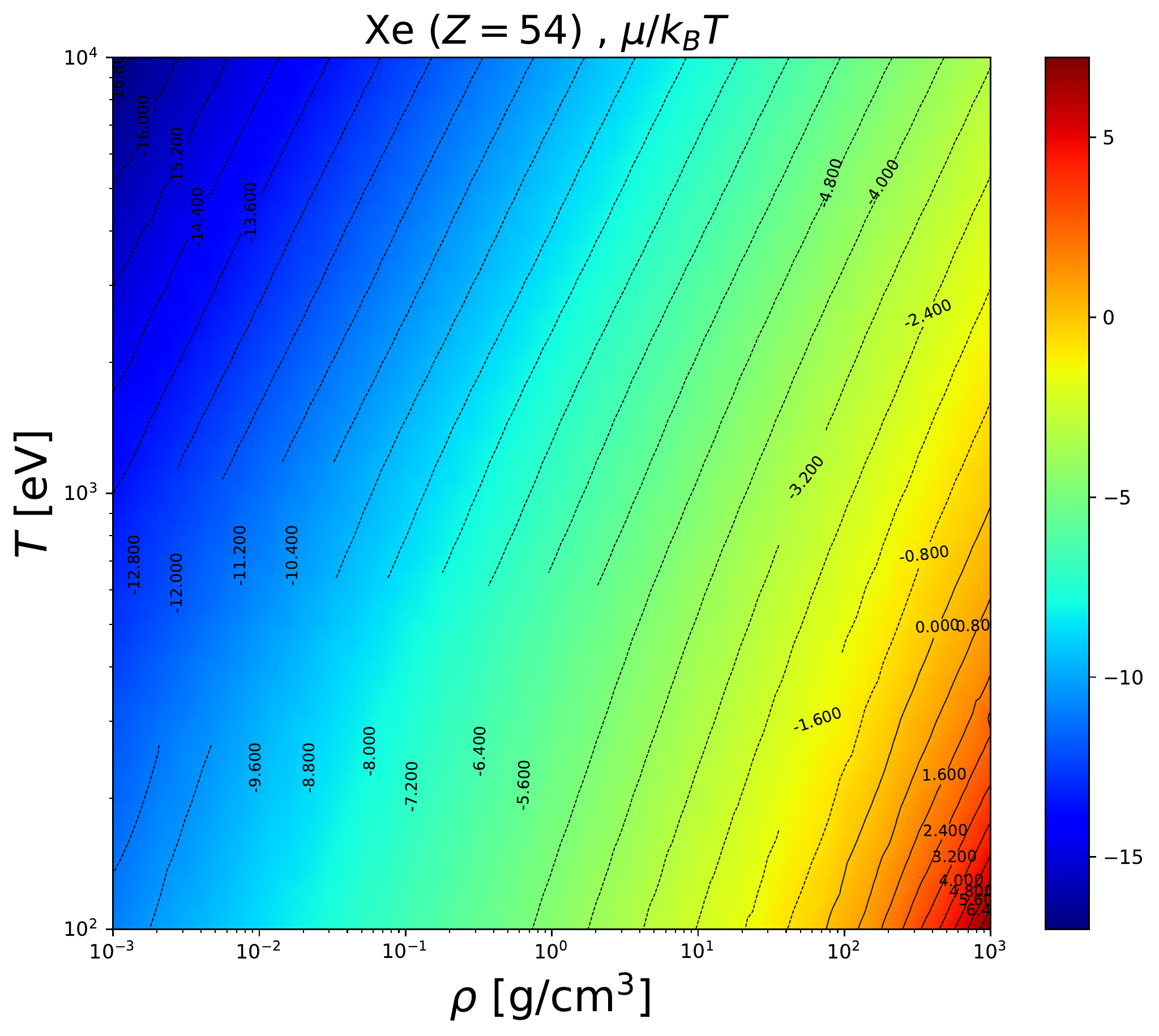} 
\par\end{centering}
\caption{(Color online) Same as Fig. \ref{fig:si_meshes}, for Xenon (Z=54).\label{fig:xe_meshes}}
\end{figure*}

\begin{figure*}
\begin{centering}
\includegraphics[scale=0.28]{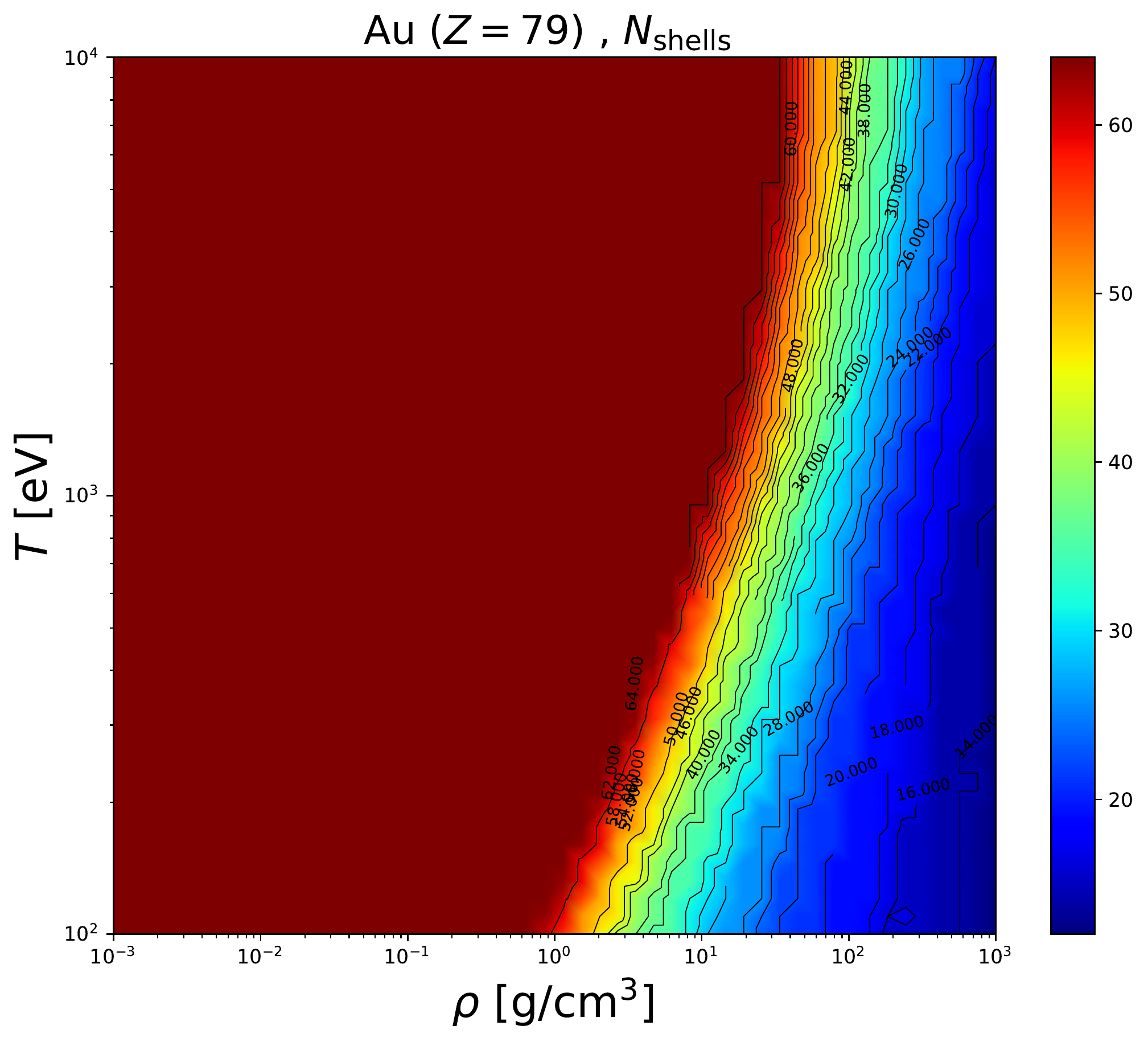}\includegraphics[scale=0.28]{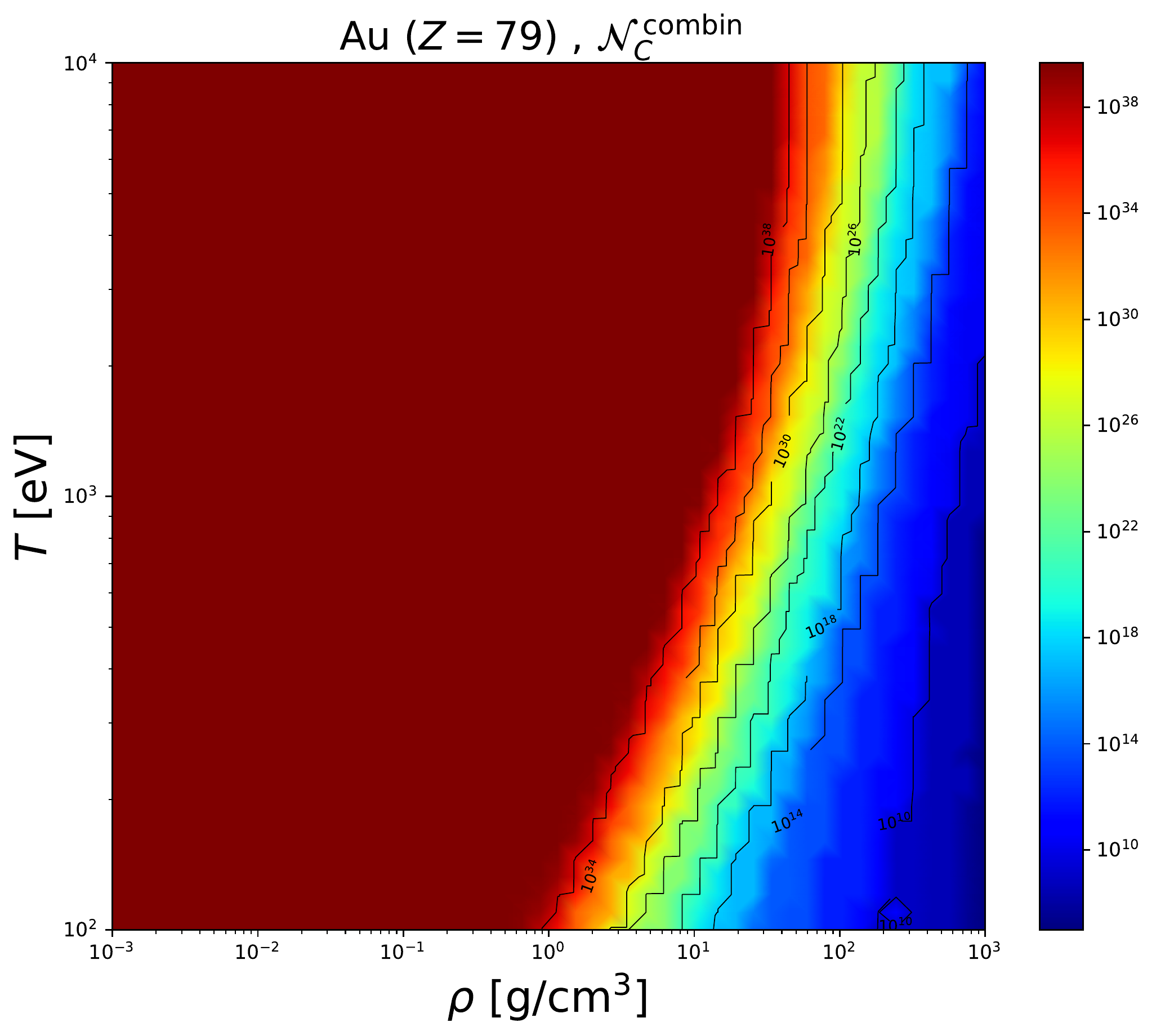}\includegraphics[scale=0.28]{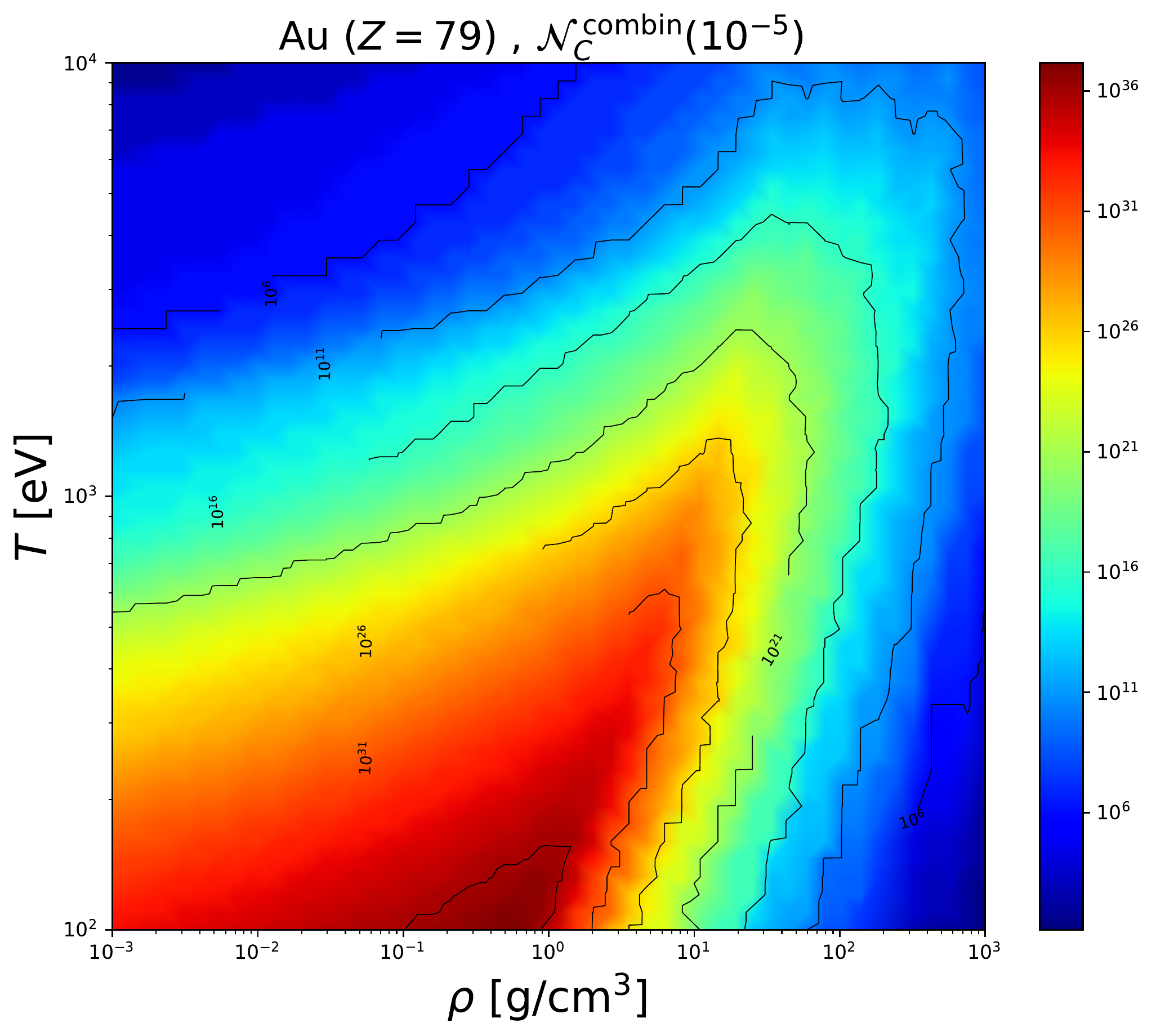}
\par\end{centering}
\begin{centering}
\includegraphics[scale=0.28]{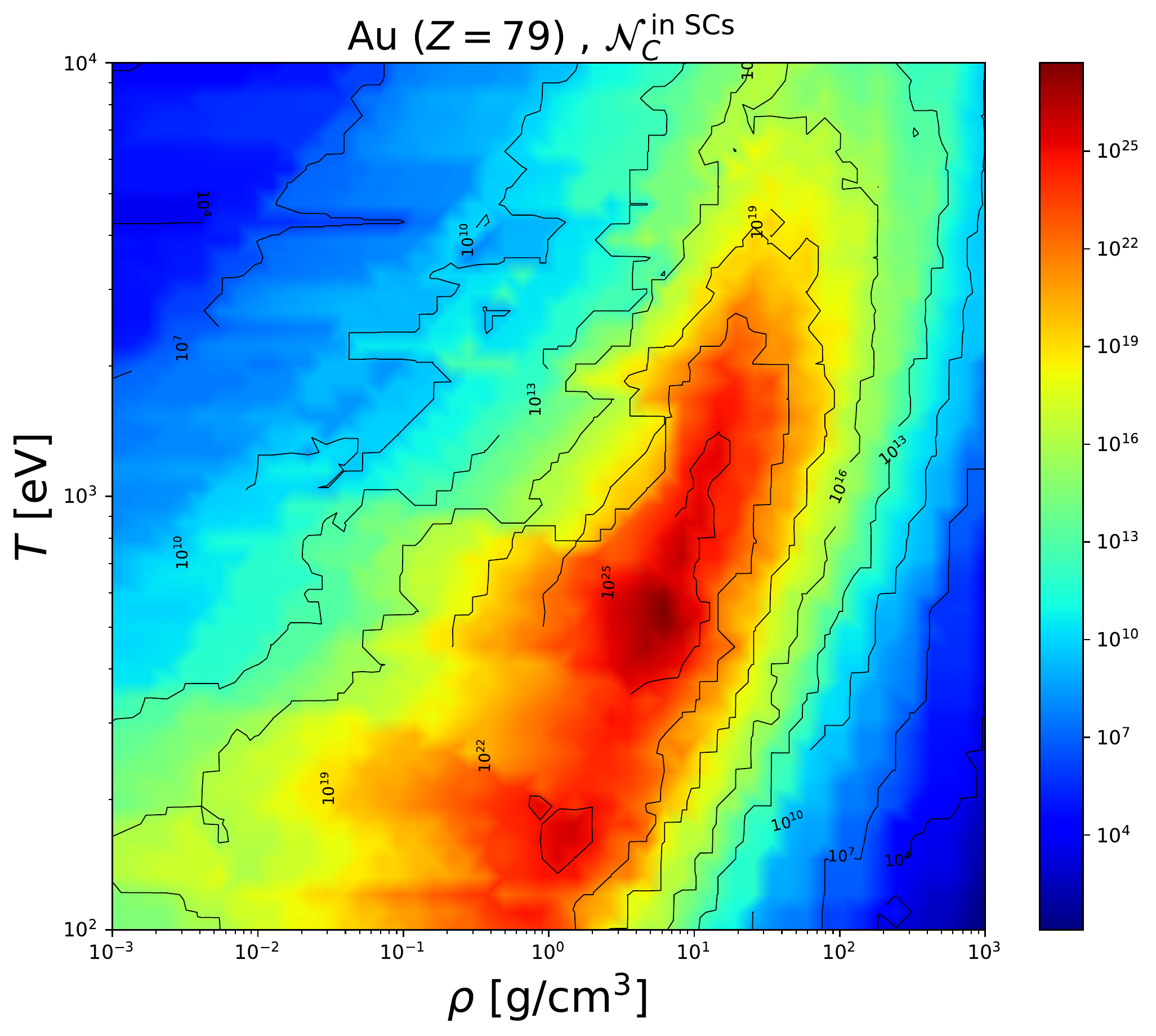}\includegraphics[scale=0.28]{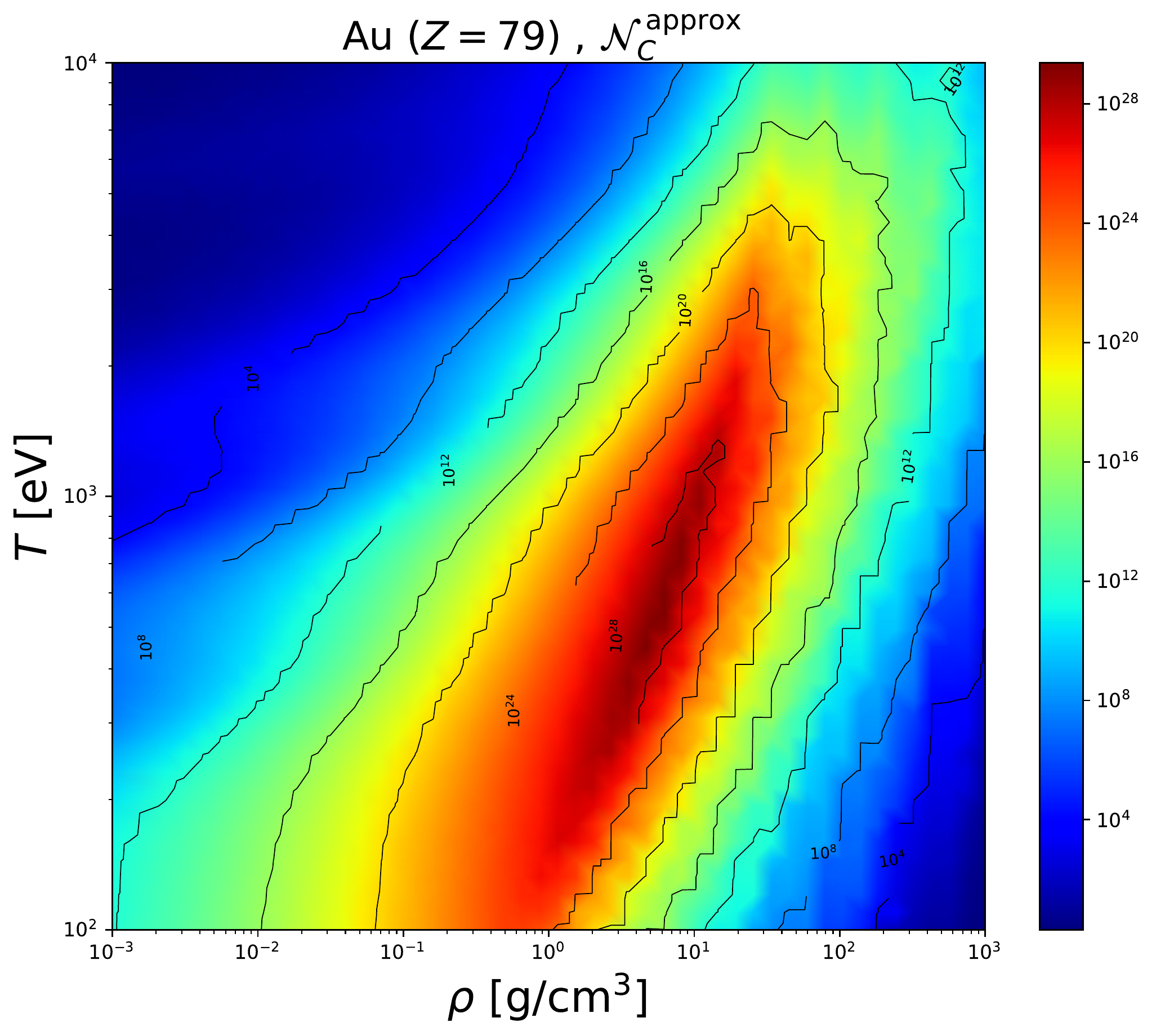}\includegraphics[scale=0.28]{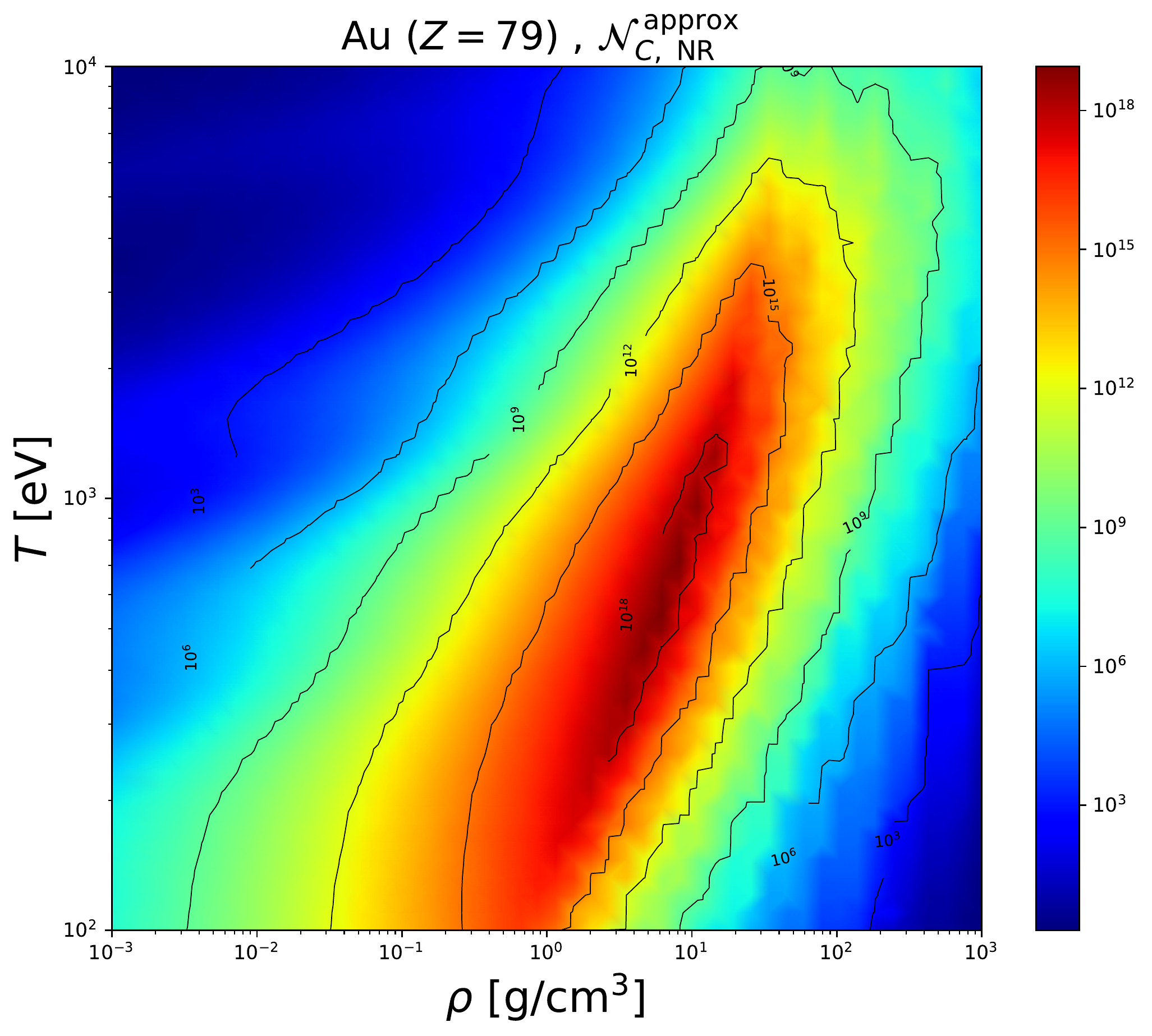} 
\par\end{centering}
\begin{centering}
\includegraphics[scale=0.28]{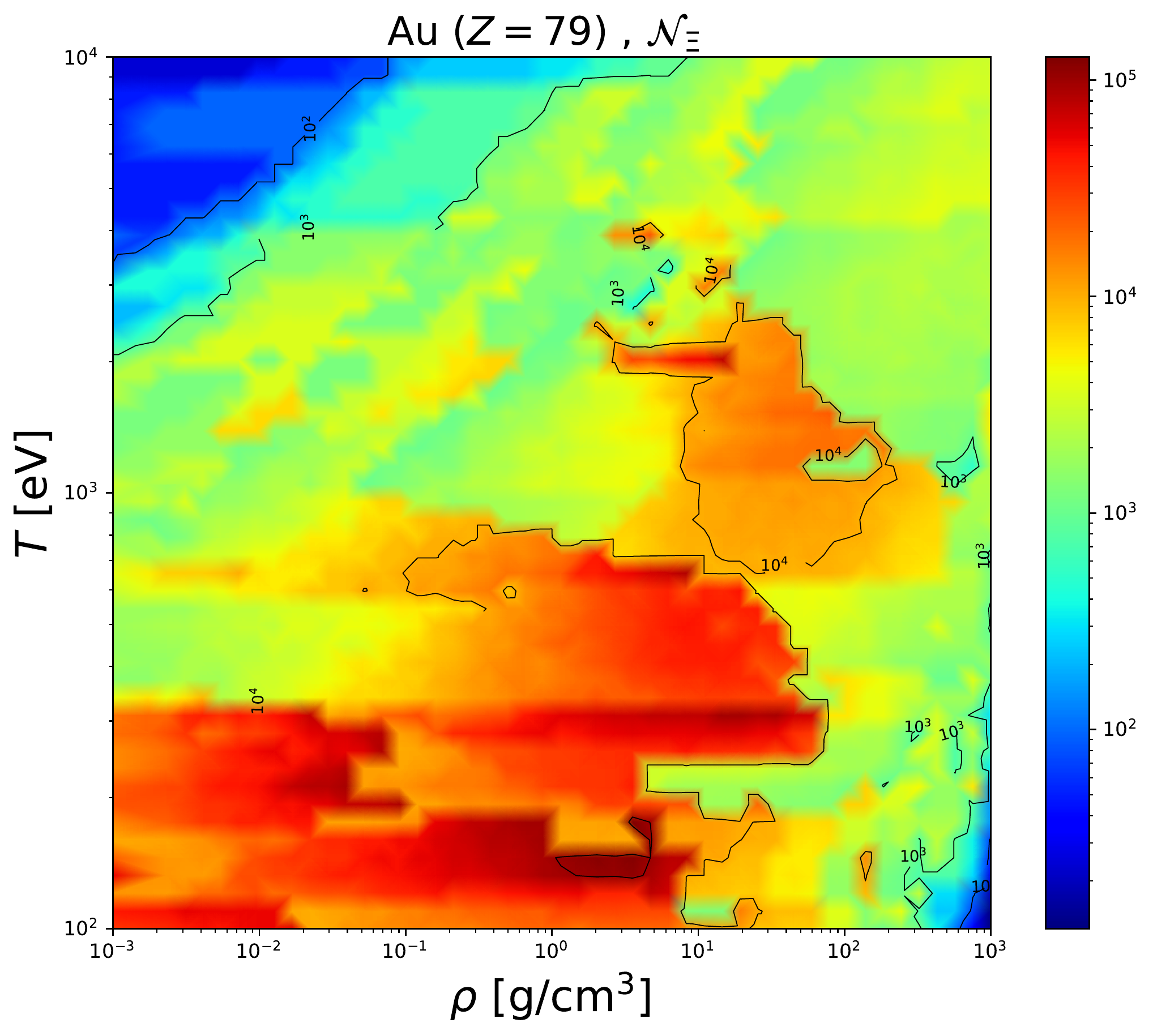}\includegraphics[scale=0.28]{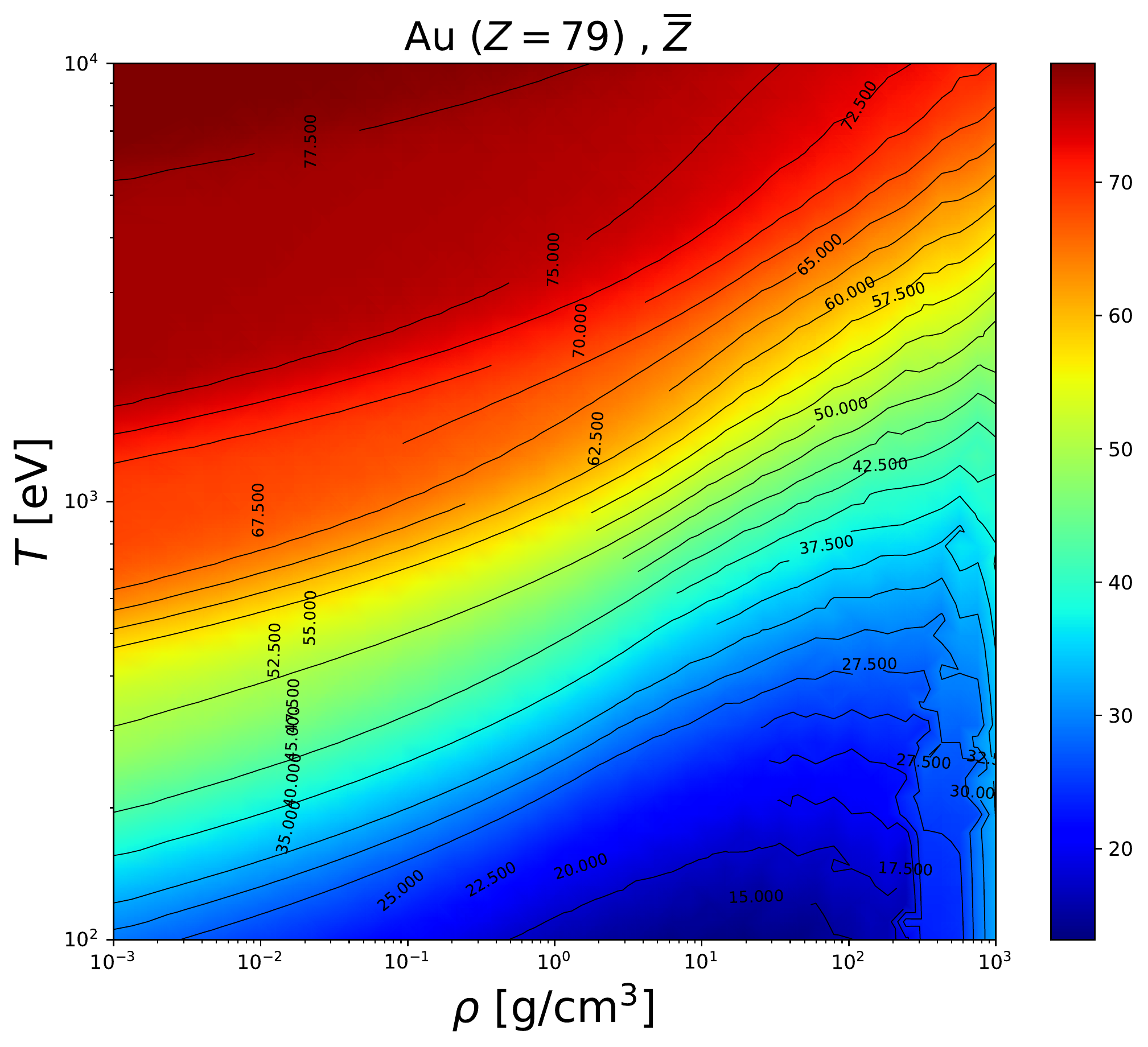}\includegraphics[scale=0.28]{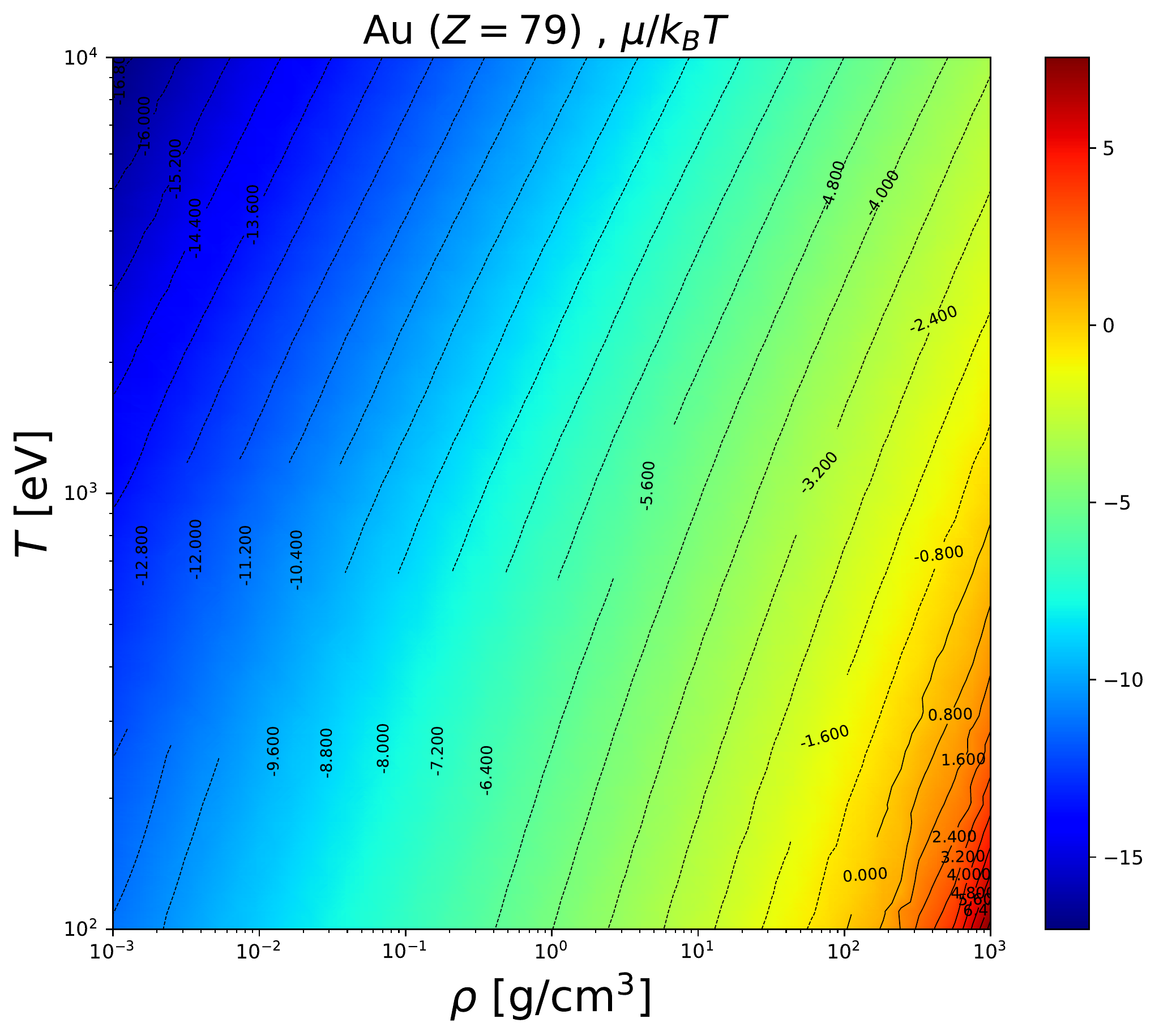} 
\par\end{centering}
\caption{(Color online) Same as Fig. \ref{fig:si_meshes}, for Gold (Z=79).\label{fig:au_meshes}}
\end{figure*}

\begin{figure}
\begin{centering}
\includegraphics[scale=0.5]{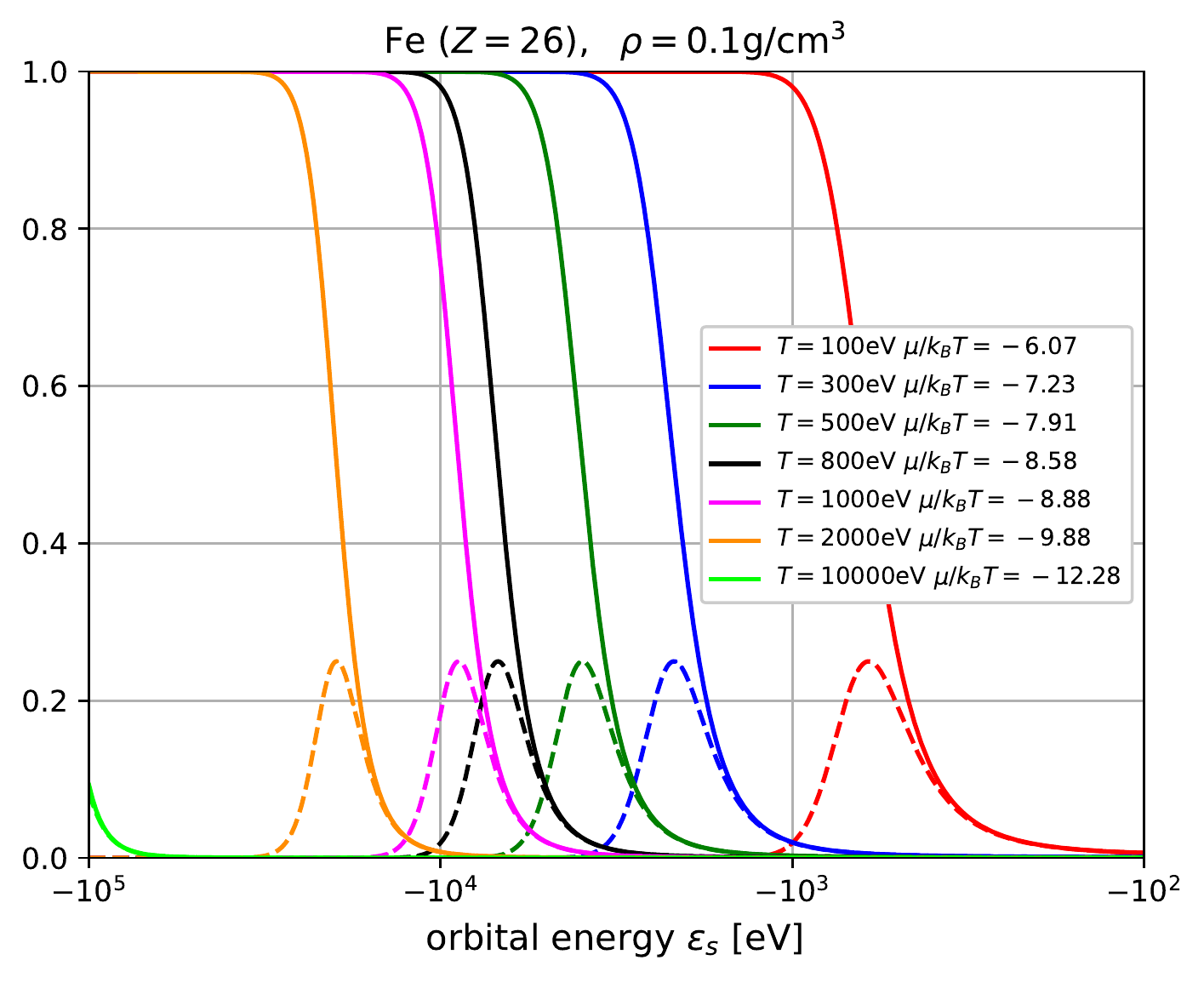} 
\par\end{centering}
\caption{(Color online) The Fermi-Dirac distribution (solid lines) and the occupation fluctuation
$\delta q_{s}^{2}/g_{s}$ (dashed lines, see eq. \ref{eq:fluc}) as
a function of orbital energy, for Iron at $\rho=0.1\text{g/cm}^{3}$,
and different temperatures in the range $10^{2}-10^{4}\text{eV}$ (lines are arranged from right to left.
\label{fig:fd_fluc_fe_T}}
\end{figure}

\begin{figure}
\begin{centering}
\includegraphics[scale=0.5]{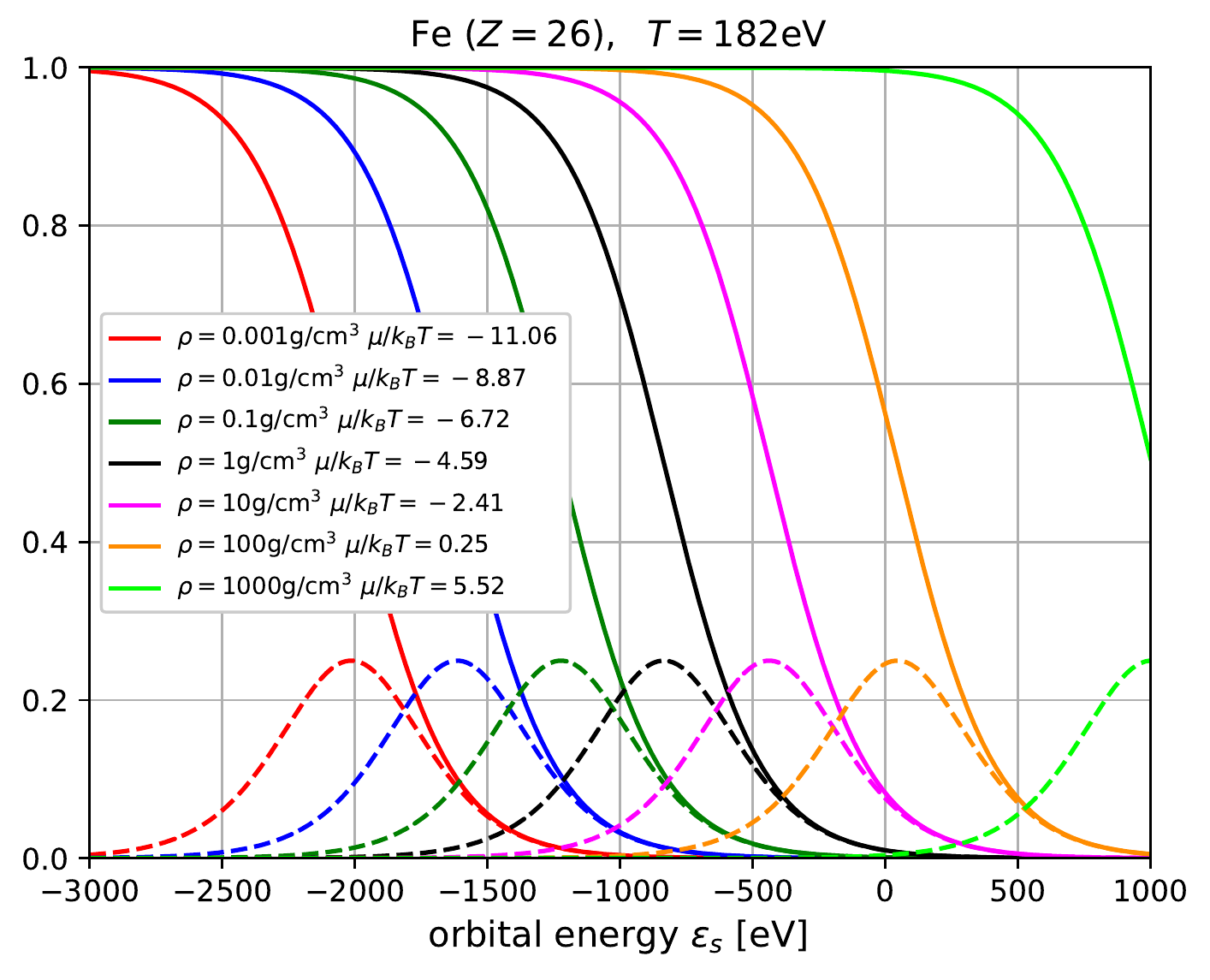} 
\par\end{centering}
\caption{(Color online) The Fermi-Dirac distribution (solid lines) and the occupation fluctuation
$\delta q_{s}^{2}/g_{s}$ (dashed lines, see eq. \ref{eq:fluc}) as
a function of orbital energy, for Iron at $T=182\text{eV}$, and different
densities in the range $10^{-3}-10^{3}\text{g/cm}^{3}$ (lines are arranged from left to right).\label{fig:fd_fluc_fe_rho}}
\end{figure}

Detailed finite temperature density functional theory (DFT) calculations
in the spherical average atom approximation \cite{liberman1979self,Rozsnyai1972,wilson2006purgatorio,blenski1995pressure,Novikov2011,ovechkin2014reseos,ovechkin2019plasma,krief_dh_2018_apj,gill2017tartarus,starrett2019wide}
followed by STA calculations using the atomic code STAR \cite{krief2016solar,krief2018new},
were performed over a wide range of plasma temperatures: 100eV-10keV
and densities: $10^{-3}-10^{3}\text{g/cm}^{3}$, for the following
low, mid and high Z elements - Silicone (Z=14), Iron (Z=26), Xenon
(Z=54) and Gold (Z=79). The results are shown in Figs \ref{fig:si_meshes}-\ref{fig:au_meshes},
and include the average ionization, the chemical potential, the number
of bound shells, the combinatoric number of configurations over all
ionization levels (eq. \ref{eq:NC_COMBIN}) and over all ionization
levels with probabilities larger than $10^{-5}$ (eq. \ref{eq:NC_COMBIN_P},
the charge distribution was obtained from the STA calculations), the
number of populated superconfigurations in a the converged STA calculation,
the number of configurations within these superconfigurations and
the approximated number of relativistic and nonrelativistic populated
configurations. First, it is evident that, as expected, the number
of configurations grows exponentially with the atomic number. It is
seen that even for mid $Z$ plasmas, the number of populated configurations
can be huge (larger than $10^{13}$) in a wide range of temperature
and density, and a detailed configuration accounting (DCA) calculation
may be extremely costly, while a full detailed-line-accounting (DLA)
calculation is probably impossible. For a higher Z plasma such as
Xenon and Gold, there is a wide range of temperature and density with
a populated number of configurations larger than $10^{20}$, which
is completely intractable for a DCA calculation, and highlights the
need for an STA \cite{BarShalom1989,blenski2000superconfiguration,hazak2012configurationally,ovechkin2014reseos,wilson2015partially,krief2016solar,krief2018star,bauche2015atomic}
or Average-Atom \cite{shalitin1984level,stein1985average,rozsnyai2001solar}
method for opacity calculations.

A comparison of the number of configurations within superconfigurations
$\mathcal{N}_{C}^{\text{in SCs}}$ and approximated number of populated
relativistic configurations $\mathcal{N}_{C}^{\text{approx}}$, shows
a good qualitative and even quantitative agreement - which proves
that $\mathcal{N}_{C}^{\text{approx}}$ is a very good simple estimate
for the number of populated configurations. In addition, the plots
for the number of superconfigurations $\mathcal{N}_{\Xi}$ in a converged
STA calculation in comparison the the number of populated configurations
highlights the strength of the STA method - a computationally tractable
number of superconfigurations (in the range of $10^{3}-10^{5}$),
which may contain a huge number of configurations (more than $10^{25}$
in some cases) - yields a converged opacity calculation, which would
have been completely prohibitive in a configuration based DCA calculation.

It is seen that the $\mu/k_{B}T$ contours are approximately straight
lines (when the density and temperatures are plotted on a log scale),
a fact which agrees with the ideal gas result: 
\begin{equation}
\mu_{\text{ideal}}=k_{B}T\ln\left(\Lambda^{3}\bar{n}\right),
\end{equation}
where $\bar{n}$ is the number density and $\Lambda=\left(2\pi\hbar^{2}/mk_{B}T\right)^{\frac{1}{2}}$
is the thermal wavelength, so that in the thermodynamic range studied
here, the chemical potential is a decreasing function of temperature
and an increasing function of density, as shown in Figs \ref{fig:fd_fluc_fe_T}-\ref{fig:fd_fluc_fe_rho}.

It is also evident from Figs \ref{fig:si_meshes}-\ref{fig:au_meshes},
that the combinatoric number of configurations over all ionization
levels correlates perfectly with the number of bound shells that exists
in the atomic potential. This is to be expected since this number
of configurations depends on temperature and density only through
the number of existing bound shells - and not on their properties
(i.e. bound energies, wave functions etc). As was explained in the
previous section, since we are concerned in the number of configurations
which should be taken into account in opacity calculations, the number
of bound shells is limited here to $64$ - which results in a sharp
front in the plots for the combinatoric number of configurations.
It is also evident that $\mathcal{N}_{C}^{\text{combin}}\left(p=10^{-5}\right)$
has a better agreement with $\mathcal{N}_{C}^{\text{in SCs}}$ and
$\mathcal{N}_{C}^{\text{approx}}$ for the lower Z elements and for
cases with a smaller number of bound orbitals. As was discussed in
the previous section, this is to be expected, since as apposed to
the superconfiguration accounting approach, the charge probability
distribution does not contain information about the configuration
structure and as a result, some charge states may contain a huge amount
of very low probability configurations, which are taken into account
in eq. \ref{eq:NC_COMBIN_P}.

Next, we discuss the temperature and density behavior for the populated
number of configurations. As expected, it is seen in Figs \ref{fig:si_meshes}-\ref{fig:au_meshes}
that the number of populated configurations has a maximum as a function
of temperature and density. For low temperatures, most shells have
energies $\epsilon_{s}<\mu$ and are therefore ``frozen'' - either
full or empty (see Fig. \ref{fig:fd_illustration}), while for moderate
temperatures (which are different for each element and density) the
Fermi-Dirac distribution has a shape of a step function, but with
a finite width which allow large fluctuations for the occupation numbers
of shells with energies nearby the step (see Fig. \ref{fig:fd_fluc_fe_T}).
For higher temperatures, on the one hand, more shells are ionized
- which results in a decrease in the number of populated configurations,
and on the other hand, the number of bound shells can be larger, due
to a wider spatial extent of the atomic central potential - which
reduces the effect of pressure ionization, as seen in the plots of
the number of bound shells. The latter effect results in the slight
tilted maxima for the number of populated configurations as a function
of temperature and density, seen in Figs \ref{fig:si_meshes}-\ref{fig:au_meshes}.
In addition, it is evident that for low densities the number of populated
configurations is small due to the decrease in the chemical potential
(see Fig. \ref{fig:fd_fluc_fe_rho}), which reduces the number of
fluctuating shells, while for very high densities most shells are
pressure-ionized (as seen in the plots of the number of bound shells)
and those which are not are occupied and frozen - which leads again
to a small number of populated configurations.

\section{Summary}

Two useful methods for the estimation of the number of populated configurations
in a hot dense plasma were studied. In the first method, an exact
calculation of the total combinatoric number of configurations within
superconfigurations in a converged super-transition-array (STA) calculation
was used. In the second method, electron exchange and correlation
effects are neglected, leading to a multivariate binomial distribution
for the electronic occupation numbers, whose multidimensional width
is an approximation for the number of populated configurations. The
mechanism which leads to the huge number of populated configurations
- namely, the fluctuations of electronic occupation numbers of bound
shells nearby the Fermi-Dirac step, is demonstrated and discussed
in detail. Comprehensive average atom finite temperature DFT calculations
are performed in a wide range of temperature and density for several
low, mid and high Z plasmas, showing a good agreement between these
two methods. In addition, the temperature and density dependence is
discussed and explained.

The second method, which is much more simple than the first, only
the bound shells and chemical potential are needed - so that the estimate
for the number of populated configurations can be obtained, for example,
by solving the Dirac equation in a Thomas-Fermi potential or in a
more advanced average atom model potential. This simple estimate can
be very useful in order to asses the computational ability to perform
configuration based, or even line based opacity and equation of state
calculations.
\begin{acknowledgments}
We thank the anonymous referees for useful suggestions and comments,
and in particular, for suggesting the examination of the number of
configurations with respect to probability thresholds and a discussion
of possible improvements of the binomial approximation.
\end{acknowledgments}

\bibliographystyle{unsrt}
\bibliography{datab}

\pagebreak
\end{document}